\documentclass[fleqn,epsfig,usenatbib]{mn2e}
\usepackage{color} 

\usepackage{times}
\usepackage{graphicx,amsmath,amssymb,subfigure}
\usepackage{lineno}

\def\simgt{\mathrel{\lower0.6ex\hbox{$\buildrel {\textstyle >}
 \over {\scriptstyle \sim}$}}}
\def\simlt{\mathrel{\lower0.6ex\hbox{$\buildrel {\textstyle <}
 \over {\scriptstyle \sim}$}}}

\newcommand{\Msolar}{\mbox{\,$\rm M_{\odot}$}}        
\newcommand{\Lsolar}{\mbox{\,$\rm L_{\odot}$}}        

\newcommand{\gtsim}{\mbox{{\raisebox{-0.4ex}{$\stackrel{>}{{\scriptstyle\sim}}
$}}}}

\title[The environmental density of far-infrared bright
galaxies]{{\textit{Herschel}}\thanks{Herschel is an ESA space
    observatory with science instruments provided by European-led
    Principal Investigator consortia and with important participation
    from NASA.}-ATLAS/GAMA: The Environmental Density of
  Far-Infrared Bright Galaxies at $z \leq 0.5$}

\author[C. S. Burton et al.]
{~C. S. Burton$^{1}$\thanks{E-mail: c.s.burton@herts.ac.uk}, 
~Matt J. Jarvis$^{2,3}$, 
~D. J. B. Smith$^{1}$, 
~D. G. Bonfield$^{1}$, 
~M. J. Hardcastle$^{1}$, 
\newauthor 
~J.A. Stevens$^{1}$, 
~N. Bourne$^{4}$, 
~M. Baes$^{5}$, 
~S. Brough$^{6}$, 
~A. Cava$^{7}$, 
~A. Cooray$^{8}$, 
~A. Dariush$^{9}$, 
\newauthor 
~G. De Zotti$^{10,11}$, 
~L. Dunne$^{12}$, 
~S. Eales$^{13}$, 
~R. Hopwood$^{9,14}$, 
~E. Ibar$^{15}$, 
~R. J. Ivison$^{15,16}$, 
\newauthor 
~J. Liske$^{17}$, 
~J. Loveday$^{18}$, 
~S. J. Maddox$^{12}$, 
~M. Negrello$^{14}$, 
~M. W. L. Smith$^{13}$ and 
~E. Valiante$^{13}$\\ 
$^{1}$Centre for Astrophysics, Science \& Technology Research
Institute, University of Hertfordshire, AL10 9AB, UK\\
$^{2}$Oxford Astrophysics, Department of Physics, Keble Road, Oxford,
OX1 3RH, UK\\
$^{3}$Physics Department, University of the Western Cape, Bellville
7535, South Africa\\
$^{4}$School of Physics \&\ Astronomy, Nottingham University,
University Park Campus, Nottingham, NG7 2RD, UK\\
$^{5}$Sterrenkundig Observatorium, Universiteit Gent, Krijgslaan 281 S9,
B-9000 Gent, Belgium\\
$^{6}$Australian Astronomical Observatory, P.O.~Box 915, North Ryde,
NSW 1670, Australia\\
$^{7}$Departamento de Astrof\'{\i}sica, Facultad de CC. F\'{\i}sicas,
Universidad Complutense de Madrid, E-28040 Madrid, Spain\\
$^{8}$Department of Physics and Astronomy, University of California,
Irvine, CA 92697, USA\\
$^{9}$Physics Department, Imperial College London, Prince Consort
Road, London, SW7 2AZ, UK\\
$^{10}$INAF - Osservatorio Astronomico di Padova, Vicolo
dell'Osservatorio 5, I-35122 Padova, Italy\\
$^{11}$SISSA, Via Bonomea 265, I-34136 Trieste, Italy\\
$^{12}$Department of Physics and Astronomy, University of Canterbury,
Private Bag 4800, Christchurch, 8140, New Zealand\\
$^{13}$School of Physics and Astronomy, Cardiff University, Queen’s
Buildings, The Parade, Cardiff CF24 3AA, UK\\
$^{14}$Department of Physics and Astronomy, The Open University,
Walton Hall, Milton Keynes, MK7 6AA, UK\\
$^{15}$UK Astronomy Technology Centre, Royal Observatory, Edinburgh,
EH9 3HJ, UK\\
$^{16}$Institute for Astronomy, University of Edinburgh, Royal
Observatory, Blackford Hill, Edinburgh EH9 3HJ, UK\\
$^{17}$European Southern Observatory, Karl-Schwarzschild-Str.~2, 85748
Garching, Germany\\
$^{18}$Astronomy Centre, University of Sussex, Falmer, Brighton BN1
9QH, UK\\
}

\begin{document}

\date{\vspace*{-3.0em}\today}


\pagerange{\pageref{firstpage}--\pageref{lastpage}} \pubyear{2012}

\maketitle


\begin{abstract}
  We compare the environmental and star formation properties of
  far-infrared detected and non--far-infrared detected galaxies out to
  $z \sim0.5$. Using optical spectroscopy and photometry from the
  Galaxy And Mass Assembly (GAMA) and Sloan Digital Sky Survey (SDSS),
  with far-infrared observations from the {\em Herschel}-ATLAS Science
  Demonstration Phase (SDP), we apply the technique of Voronoi
  Tessellations to analyse the environmental densities of individual
  galaxies. Applying statistical analyses to colour, $r-$band
  magnitude and redshift-matched samples, we show there to be a
  significant difference at the 3.5$\sigma$ level between the
  normalized environmental densities of these two populations. This is
  such that infrared emission (a tracer of star formation activity)
  favours underdense regions compared to those inhabited by
  exclusively optically observed galaxies selected to be of the same
  $r-$band magnitude, colour and redshift. Thus more highly
  star-forming galaxies are found to reside in the most underdense
  environments, confirming previous studies that have proposed such a
  correlation. However, the degeneracy between redshift and
  far-infrared luminosity in our flux-density limited sample means
  that we are unable to make a stronger statement in this respect. We
  then apply our method to synthetic light cones generated from
  semi-analytic models, finding that over the whole redshift
  distribution the same correlations between star-formation rate and
  environmental density are found.
\end{abstract}
\begin{keywords}
 method: data analysis --  galaxies: statistics -- galaxies: submillimetre-- galaxies:
  star formation.
\end{keywords}


\section{Introduction}\label{sec:Introduction}

\subsection{The environmental influence on galaxy formation and evolution}
 
It is clear that if we are to understand the process by which galaxies
form and evolve, we have to consider the role that their immediate
environment plays. \citet{Dressler1980} was the first to show that
there is a correlation between the morphology of a galaxy population
and the density of its environment. Further studies have since shown
that disk dominated `late-type' galaxy morphologies with high star
formation rates (SFR) dominate underdense regions, while their
elliptical `early-type' counterparts, with low SFR, dominate the
densest regions (\citealt{Postman&Geller1984};
\citealt{Dressler_et_al1997}; \citealt{Dominguez_et_al2001};
\citealt{Goto_et_al2003}; \citealt{Kauffmann_et_al2004};
\citealt{O'Mill_et_al2008}; \citealt{Lee_et_al2010};
\citealt{Wijesinghe_et_al2012}).

It has also been shown, with the advent of large area surveys such as
the Sloan Digital Sky Survey (SDSS; \citealt{York_et_al2000}), that
these galaxies can be categorised into two distinct optical colour
populations, `{\em Red}' and `{\em Blue}', where the colour of a
galaxy is dependent on several internal properties that represent the
evolutionary history of the galaxy; metallicity ($Z$), star formation
history (SFH) and dust attenuation ($A$). These two colour populations
show, at a fixed luminosity, a correlation with density such that the
densest regions are populated by the red, early-type passive galaxies,
with blue, star-forming late-types observed in less dense regions
\citep{Poggianti_et_al2006}. However, earlier work by
\citet{Balogh_et_al1997, Balogh_et_al1998} compared galaxies with
similar luminosities and morphologies from both dense cluster and low
density field environments, and found that that the SFR was still
lower in dense cluster regions and thus the SFR-density correlation
still held regardless of the morphology of the galaxy. This indicates
that the observed SFR-density relation cannot be exclusively tied to
the morphology-density relation; other processes must be influencing
the observed correlations. It is currently believed that this
reduction in SFR with increased environmental density is directly
linked to the stripping of cold gas from galaxies via some type of
direct interaction, and several mechanisms have been invoked to
explain this observed correlation.

For example, major mergers \citep{Barnes&Hernquist1992} can cause a
burst of star-formation activity and feedback from such star-burst
events is able to prevent gas cooling, and as a result the gas remains
out of pressure equilibrium with its environment. Due to this pressure
difference the gas expands out of the central regions of the galaxy
sweeping up the inter-stellar medium (ISM). This ejection of the ISM
from the merger remnant can lead to further suppression of star
formation (\citealt{MacLow&Ferrara1999}; \citealt{Gay_et_al2010}).
Alternatively, other processes such as harassment, strangulation and
ram-pressure stripping may also play an important role \citep[see
e.g.][for a review]{BoselliGavazzi2006}.

\subsection{Far-infrared emission as a tracer of star formation}\label{sec:Our}

Star formation within a galaxy typically increases the dust content of
the ISM through processes associated with the short-lived massive
stars that inhabit these regions, such as supernovae, that
redistribute material into the surrounding ISM
(\citealt{Dunne_et_al2003}; \citealt{Sugerman_et_al2006};
\citealt{Dunne_et_al2009}; \citealt{Gomez_et_al2012}). This dust then
absorbs a significant fraction of the ultra-violet (UV) light emitted
by the young O-B type stars associated with these regions and is
heated to temperatures of around 20-40\,K, emitting thermal radiation
at far-infrared wavelengths. This makes the use of far-infrared
emission from a galaxy a widely used diagnostic for the obscured SFR
of a galaxy (\citealt{Kennicutt1998}; \citealt{Hirashita_et_al2003};
\citealt{Driver_et_al2007}; \citealt{Cortese_et_al2008};
\citealt{Nordon_et_al2010}; \citealt{Buat_et_al2010};
\citealt{Dunne_et_al2011}; \citealt{Smith_et_al2012}).

However, other contributions to the UV radiation field which heats the
dust, such as AGN and older stellar populations within the galaxy, may
lead to overestimates of the SFR using far-infrared emission
(\citealt{Schmitt_et_al2006}; \citealt{dacunha_et_al2008};
\citealt{Nardini_et_al2008}; \citealt{Bendo_et_al2010,
  Bendo_et_al2012}; \citealt{Groves_et_al2012};
\citealt{Smith_et_al2012}; \citealt{Smith_MWL_et_al2012}). Conversely,
in galaxies where the ISM is optically thin at UV wavelengths, the
measured SFR will emerge directly from the UV and not in the
far-infrared. In these galaxies, deriving the star formation from far-infrared
emission may lead to an underestimation of the total SFR
\citep{Kennicutt1998}. However, as more than $50$\,per cent of energy
ever radiated from stars has been absorbed by dust and re-radiated
into the infrared (\citealt{Puget_et_al1996}; \citealt{Fixsen_et_al1998};
\citealt{Adelberger&Steidel2000}; \citealt{Finn_et_al2010}), with the
bulk of star formation since $z=1$ occurring in dust obscured galaxies
(\citealt{Calzetti&Heckman1999}; \citealt{Le-Floch_et_al2005};
\citealt{Patel_et_al2013}), only AGN, low metallicity systems and very
passive but dusty galaxies will lie off the far-infrared to SFR relation.

Initial studies of the relationship between SFR and infrared emission
from galaxies focused on shorter wavelengths using the {\em{Infrared
    Astronomical Satellite}} (IRAS; \citealt{Neugebauer_et_al1984})
and more recently the {\em{Spitzer Space Observatory}}
\citep{Rieke_et_al2004}. {\em IRAS} surveyed the vast majority of the
sky between $12-100$ $\mu$m, providing a large census of dusty
galaxies in the local Universe.
Using these data, \citet{Goto2005} investigated the optical properties
of 4248 infrared-selected galaxies by positionally matching data from
the IRAS with optical data from the SDSS. Using a volume limited
sample at $z \le 0.06$ and applying a $5$th-nearest neighbour density
estimate, a trend was found such that galaxies with the highest
infrared luminosities reside in relatively low-density local
environments, suggesting that star-forming galaxies favour underdense
regions, in agreement with previous studies at other wavelengths.

The environmental densities of {\em IRAS}-detected luminous infrared
galaxies (LIRGs; $10^{11} \leq L_{FIR}<10^{12}\Lsolar$) at $0.03 \le z
< 0.17$ were also studied by \citet{Hwang_et_al2010a}. They found that
the fraction of LIRGs was strongly dependent on both the morphology
and the distance to the nearest neighbour galaxy. They conclude that
the evolution of the SFR-density relation from high to low redshifts
is consistent with the idea that galaxy-galaxy interactions and
merging play a critical role in triggering star formation in LIRGs.

Additionally, \citet{Tekola_et_al2012} examined the relationship
between star formation and the environments of LIRGs selected from
{\em IRAS} and compared these with other types of high- and low-
redshift galaxies out to $z \sim 1$. They identified that there was a
luminosity ($L_{IR}\sim10^{11}h^{-2}\Lsolar$) at which infrared
selected galaxies preferentially resided in higher density
environments, compared to ``normal'' galaxies. Above this luminosity
the average density increases, whereas below this luminosity,
infrared-selected galaxies reside in environments of equal density,
similar to the general population. They conclude, therefore, that
infrared activity for non-LIRGs is not dependent on density and that
the SFR-density relationship for these galaxies is similar to that of
blue galaxies at $z \sim 1$.

At higher redshifts, \citet{Feruglio_et_al2010} used $24\mu$m
observations from {\em{Spitzer}} to investigate the environmental
effects on star formation in LIRGs and ultra-luminous infrared
galaxies (ULIRGs; $> 10^{12}\Lsolar$) in the Cosmic Evolution Survey
(COSMOS; \citealt{Scovelli_et_al2007}) at $0.3 < z < 1.2$. They found
the fraction of these galaxies to decrease with density out to $z \sim
1$, but that the relationship flattens out with increasing redshift.

Due to the wavelength coverage of IRAS ($12-100\mu$m), the majority of
galaxies detected by these studies were found by
\citet{Bregman_et_al1998} to be spirals and starbursts in the local
universe ($z < 0.1$). This restriction resulted in the IRAS providing
little information about the cooler dust, which traces the bulk of the
dust mass \cite[e.g.][]{Dunne_et_al2011}, in other galaxy populations,
especially early-type morphologies. In comparison, {\em{Spitzer}} can
probe longer wavelengths ($24-160\mu$m) and therefore is less
susceptible to this bias, although galaxies with the coldest dust
temperatures would still be missed (\citealt{Eales_et_al2010};
\citealt{Symeonidis_et_al2011, Symeonidis_et_al2013}). Considering
that cold dust is present across all types of galaxies and is a major
contributor to infrared luminosity \citep{Willmer_et_al2009}, and
closely traces the total dust mass, it is crucial that we are able to
select galaxies at longer wavelengths.

With the launch of the {\em Herschel Space Observatory}
\citep{Pilbratt_et_al2010} we are now able to select galaxies at these
longer wavelengths. A number of studies have begun to investigate how
star formation in a galaxy, traced by far-infrared emission {at $\ge
  250\mu$m, is linked to the environment in which the galaxy
  resides. \citet{Dariush_et_al2011} used far-infrared data from the
  {\it{Herschel}} Astrophysical Terahertz Large Area Survey (H-ATLAS;
  \citealt{Eales_et_al2010}) Science Demonstration Phase (SDP) to
  examine the ultraviolet and optical properties and environments of
  low redshift galaxies ($0.02 \leq z \leq 0.2$) from the SDSS and the
  Galaxy And Mass Assembly survey (GAMA; \citealt{Driver_et_al2011};
  \citealt{Hill_et_al2011}; \citealt{Baldry_et_al2010}). They found
  that H-ATLAS detects predominantly blue/star-forming galaxies, with
  a minor contribution from 
  red galaxies (comprising highly obscured and passive systems). Using
  the 5th-nearest neighbour as an estimate of the environmental
  density, they found that the fraction of H-ATLAS detected galaxies
  is much higher ($\sim 70$\,per cent) in low-density environments
  compared to high-density environments, where the fraction was found
  to be $\sim 30$\,per cent. However, the detection rate of red and
  blue galaxies appears to be similar for both high- and low-density
  environments, indicating that it is the colour of a galaxy, rather
  than the density of its local environment, that governs whether it
  is detectable by H-ATLAS.

A consistent result was also found by \cite{Rowlands_et_al2012}, who
found that H-ATLAS detected early-type galaxies tend to have bluer
(NUV-r) colours, higher SSFRs and younger stellar populations than
optically observed early-type morphologies. They compare 354 spiral
and 30 early-type galaxy morphologies at low redshift ($z < 0.18$),
finding no significant difference between the environmental densities
of these populations. However, it is possible that they are not
sampling a large enough range of environments with such small
population samples.

\citet{Coppin_et_al2011} also used far-infrared data from H-ATLAS to examine
the centres of $66$ optically selected galaxy clusters at $z\sim0.25$
to search for statistical evidence of obscured star formation in the
cluster population. Using Voronoi Tessellations (described in Section
\ref{sec:Voronoi Tesselation Code}) to locate cluster members, they
found an excess in the surface density of far-infrared sources within
$\sim1.2$Mpc of the centre of these clusters. They conclude that the
far-infrared emission is associated with dust-obscured star formation in
cluster galaxies, translating to a rate of $\sim7 \Msolar
$\,yr$^{-1}$. This SFR, maintained over the 3 Gyr since $z=0.25$,
would contribute enough mass to construct a typical S0-type bulge that
would match the observed increase in bulge-dominated galaxies in cores
of clusters over the same timescale.

The effects of environment on the far-infrared properties of galaxies
are also discussed by \citet{Davies_et_al2010}, who use data from the
{\it{Herschel}} Virgo Cluster Survey (HeViCS;
\citealt{Davies_et_al2012}), finding that relatively few faint
far-infrared sources that can be associated with confirmed Virgo
cluster members. Furthermore, studies by
\cite{Cortese_et_al2010a,Cortese_et_al2010b} present {\it{Herschel}}
observations of the perturbed galaxy NGC $4438$ in the Virgo cluster
and identify regions of extra-planar dust up to $\sim 4-5$ kpc away
from the galaxy disk. This dust is found to closely follow the
distribution of stripped atomic and molecular hydrogen, supporting the
idea that gas and dust are perturbed in a similar way within the
cluster environment.

In contrast to these results, \citet{Geach_et_al2011}, using 24$\mu$m
observations from
{\em{Spitzer}}, investigated large-scale filamentary structure
surrounding rich clusters out to $z \sim 0.55$, and found that the
SFRs of individual galaxy members within a cluster are not
significantly different to identically selected field
galaxies. Although pockets of enhanced star formation were observed,
they suggest that this is the result of some `pre-processing' effect
where satellite groups have star formation triggered via gravitational
tidal interactions during cluster infall. However, they state that
there is no environmental mechanism acting to enhance the star
formation within individual galaxies.

It is evident that the majority of these studies have either used
density measures that do not detect differences on the smallest scales
(i.e. $n$th-nearest neighbour or aperture gridding) and/or they have
focused entirely on narrow and local redshifts ($z \lesssim 0.2$). In
this paper we use data from H-ATLAS to investigate the environmental
dependence of far-infrared emission using a technique based on Voronoi
Tessellations. Unlike the $n$th-nearest neighbour technique, Voronoi
Tessellations calculate the environmental density of galaxies on
individual galaxy scales and hence can probe the environmental density
to a greater degree of accuracy.



In Section~\ref{sec:Data} we outline the optical and infrared data that we
use. In Section~\ref{sec:METHOD} we present how both the spectroscopic
and photometric redshifts for our sample of galaxies were measured and
sampled
and introduce our algorithm to estimate the environmental density. In
Section~\ref{sec:Analysis} we present the results of our analysis to
determine whether there are any differences in environmental density
between the far-infrared bright and faint sources, and investigate
whether the SFR is linked to the environmental density. In
Section~\ref{sec:MillSim} we compare our results to semi-analytic
models and discuss the physical mechanisms that may explain our
results. In Section~\ref{sec:Discussion} we discuss our results in the
context of the physical mechanisms outlined above and in
Section~\ref{sec:Conclusions} we summarise our findings. We adopt a
cosmology throughout with $\Omega_{m} = 0.30$, $\Omega_{\Lambda} =
0.70$ and $H_{0} = 71 $ km{\rm s}$^{-1}$ Mpc$^{-1}$.

\section{Observations}\label{sec:Data}

\subsection{Far-infrared data}\label{sec:IR Data}

We use far-infrared data from the science demonstration phase of
H-ATLAS (\citealt{Rigby_et_al2011}). H-ATLAS provides data across a
wavelength range of 100-500 $\micron$ using the Photo-detector Array
Camera and Spectrometer (PACS; \citealt{Poglitsch_et_al2010}) at 100
and 160 $\micron$; and the Spectral and Photometric Imaging REceiver
(SPIRE; \citealt{Griffin_et_al2010}) at 250, 350 and 500$\micron$. The
H-ATLAS observations consist of two scans in parallel mode reaching
5$\sigma$ point source sensitivities of 132, 126, 32, 36 and 45 mJy in
the 100, 160, 250, 350, and 500$\mu$m channels, respectively, with
beam sizes of approximately 9, 13, 18, 25 and 35 arcsec in the same
five bands. The SPIRE and PACS map-making procedures are described by
\citet{Pascale_et_al2011} and \citet{Ibar_et_al2010} respectively,
while the catalogues are described by
\citet{Rigby_et_al2011}. \citet{Smith_et_al2011} used a likelihood
ratio (LR) method to associate optical counterparts with the H-ATLAS
galaxies down to a limiting magnitude of $r = 22.4$ within a 10 arcsec
radius. This resulted in optical counterparts for $2,423$ objects from
the H-ATLAS 250$\mu$m catalogue, each with a reliability $R>0.8$ which
ensures not only that the contamination rate is low but also that only
one $r-$band source dominates the far-infrared emission. While the
entire H-ATLAS survey aims to compile a catalogue of $\sim$10$^{5}$
extra-galactic far-infrared sources out to z $\sim$ 3, the SDP field
covered $\sim$ 3 per cent of this, over an area of $\sim$ 4.0 deg
$\times$ 3.6 deg centred on (09$^{h}$05$^{m}$, +0$^{\circ}$30'). To
maintain consistency with our Optical catalogue outlined in Section
\ref{sec:Optical Data}, we limit our far-infrared catalogue magnitude
to $r=21.5$.

\subsection{Optical and near-infrared data}\label{sec:Optical Data}

We use spectroscopic redshifts from both the SDSS and the GAMA survey
Data Release One (DR1). Spectroscopic redshifts are provided for
magnitude limits of $r < 19.4$ , $K < 17.6$ and $z < 18.2$ in the GAMA
9hr (G09) field which includes the H-ATLAS SDP. This is combined with
photometric redshifts derived from the combination of optical
({\em{ugriz}}) SDSS and near-infrared ({\em{YJHK}}) UKIDSS-LAS imaging
data 
as detailed in \citet{Smith_et_al2011}. This complete
optical--near-infrared catalogue, containing photometric and
spectroscopic redshifts (hereafter named the Optical-9hr catalogue),
totals 909,985 objects from which we remove all sources with $r >
21.5$ due to the fact that at fainter magnitudes the signal-to-noise
ratio deceases to an extent where errors associated with the
photometry become large. In addition, we remove objects classified as
point-like in the SDSS imaging. This reduced the Optical-9hr catalogue
to $323,969$ objects, of which $8,875$ had spectroscopic redshifts
across a redshift range of $0 < z < 1.2$.

\section{Environmental Density Measurement}\label{sec:METHOD}

In this section we describe our method of determining the
environmental density of individual galaxies in redshift
slices. First, as the vast majority of the galaxies in our sample do
not have spectroscopic redshifts, we are forced to use their
photometric redshifts in order to establish where they reside in three
dimensional space. Spectroscopic redshifts have errors of the order of
$\Delta z \sim 10^{-4}$ \citep{Driver_et_al2011} with the average
error associated with our photometric redshifts of the order of
$\Delta z \sim 0.16$. As these photometric redshifts apply to both
H-ATLAS and non--H-ATLAS sources, both populations would experience
similar biases associated with these errors. Adopting a
single redshift at the peak of the photometric redshift probability
distribution (z-PDF) would not accurately represent our limited
knowledge of the redshift of individual galaxies. We therefore use the
full photometric z-PDF to carry out Monte-Carlo (MC) simulations which
sample each z-PDF 1000 times generating 1000 MC cubes for each
object. From these samples we ensure that we have a good statistical
representation of the 3-dimensional distribution of galaxies within
the survey area. However, where available, we use spectroscopic
redshifts due to their smaller uncertainties.


\subsection{Voronoi Tessellations}\label{sec:Voronoi Tesselation Code}

\begin{figure}
\centering
\includegraphics[width=1.0\columnwidth]{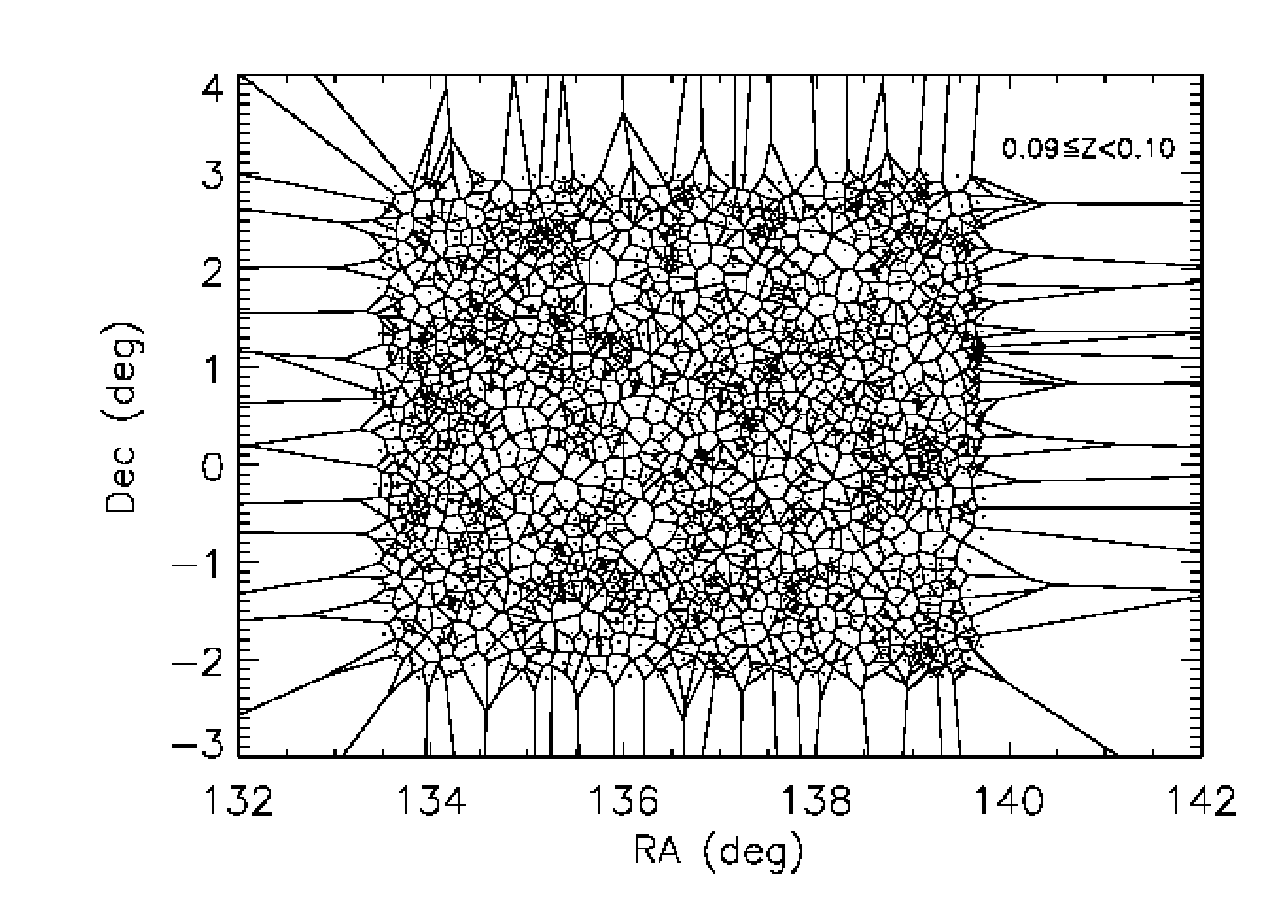}
\caption{An example of how the VT algorithm works,
  showing a subset of objects placed within the redshift slice 0.09
  $\leq$ z $<$ 0.10. The VT cells are plotted illustrating where the
  densest regions lie. As the algorithm requires more than one object
  within a specific region to accurately construct a cell boundary,
  the lack of neighbours around the edge of the image results in
  overly large cell areas. As explained in the main text (Section
  \ref{sec:Voronoi Tesselation Code}), these outer cells are removed
  from the rest of the analysis.}
\label{fig:VT2}
\end{figure} 

In order to calculate the environmental density of individual galaxies
we apply a numerical algorithm called `Voronoi Tessellations' (VT;
\citealt{Icke&van-de-Weygaert1987};
\citealt{van_de_Weygaert&Icke1989}) to the Optical-9hr catalogue. The
algorithm works by initially treating each object in the field as a
single point source (or nucleus). It then constructs a convex polygon
(or `Voronoi cell') around each of these nuclei enclosing all points
that are closer to that nucleus than any other.
The area of the Voronoi cell is a good representation of the local
environment of that object, such that the reciprocal of this area
gives a direct measure of the density. The VT technique has been used
in many areas of astronomy, initially in the study of large-scale
structure of the universe (e.g., \citealt{Icke&van-de-Weygaert1987};
\citealt{van_de_Weygaert&Icke1989}; \citealt{Diehl&Statler2006}) and
more recently in studies of cluster detection
(e.g., \citealt{Kim_et_al2000}; \citealt{Ramella_et_al2001};
\citealt{van-Breukelen_et_al2006}; \citealt{van-Breukelen_et_al2006b};
\citealt{van-Breukelen_et_al2009}; \citealt{Geach_et_al2011b};
\citealt{Soares-Santos_et_al2011}).

The VT algorithm does not take redshift into account when linking each
nucleus together as we lack the necessary resolution in redshift to
apply a three-dimensional VT algorithm. Therefore it is necessary to
group galaxies into redshift slices so as to avoid projection
effects. Subsequently each of the 1000 3D MC fields are split into
redshift slices so that the VT algorithm can be applied to each slice
individually; thus only objects within each slice have the VT
algorithm applied to them, maximising the accuracy of the density
calculation for each object. The width of each slice is limited such
that, if too wide, projection effects may become a problem. In
addition, if the slice is too small, an overdense region in terms of
the rest of the field may go undetected by the VT algorithm if it is
spread across multiple slices. The typical velocity dispersion between
gravitationally bound group/cluster galaxies is within the region of a
few hundred kilometres per second (\citealt{Haynes_et_al1984};
\citealt{Martini_et_al2007}),
which equates to a spread in redshift of $\Delta z/(1+z) \sim 10^{-3}$
at $z =
1$. 
Therefore we adopt a redshift slice of width $\Delta z = 0.01$, easily
incorporating associated galactic environments and resulting in 120
slices across each of the 1000 3D MC realisations across the full
redshift range of our data out to $z= 1.2$.

For each object in each MC cube realisation its Voronoi cell area
($x_{i}$) is calculated, the mean of which ($\bar{x_{i}}$) gives the
overall mean area calculated for that object across all of the MC
realisations. Taking an inverse of this mean area gives a value for
the mean environmental density ($\bar\rho_{i}$) for that
object. 
Figure~\ref{fig:VT2} shows an example of one VT slice (containing a
smaller subset of the data for illustrative purposes) and it is
immediately clear that objects towards the outside of the field have
overly large Voronoi cell areas. This edge effect is the result of the
VT algorithm not finding any objects outside of this boundary and
consequently being unable to triangulate in these areas. In order to
prevent this edge effect altering the mean density result, a cut is
then made around the outside of the field to remove the outermost
objects (and their overly large cell areas) from the Optical-9hr
catalogue. The position of this cut was calculated by plotting the
right ascension (RA) and declination (Dec) values separately against a
value that represents a normalised value for the mean area, the
significance ($S$) (this value is introduced to account for a peak in
the number density of objects as explained in Section \ref{sec:Density
  Normalization}). The resultant plots showed, unsurprisingly, a sharp
increase in mean area of the cells towards the outside of the field
and that this edge effect penetrated the field by approximately $\pm
0.30$~degrees in both RA and Dec. As the Optical-9hr field extends
well beyond the H--ATLAS-SDP field on all sides, this cut is not
significant in terms of the number of sources lost and does not
interfere with the accuracy of our analysis.


Furthermore, to ensure that the accuracy of the comparison between the
two samples is maintained, it is necessary to include only Optical-9hr
objects that reside within the boundary of the H--ATLAS-SDP
field. This ensures that all objects included in the density measure
are from across the same region and thus have been observed by both
SDSS and H-ATLAS observations. Thus when comparing far-infrared
detected and undetected galaxies we are not counting any far-infrared
luminous galaxies that would otherwise be detected in the H-ATLAS SDP
catalogue if it were not for the boundary limits of the H-ATLAS SDP
region.
After these region cuts are applied, the final Optical-9hr catalogue
is reduced to 129,518 objects. In section \ref{sec:KS-testing0}, this
catalogue is divided according to whether or not the galaxies have
far-infrared emission in order that these sub-samples can then be
compared. Table \ref{tab:Optical-9hr_Breakdown} shows the number of
galaxies within these sub-samples in addition to the number of
galaxies with photometric and spectroscopic redshifts.


\subsection{Density normalization}\label{sec:Density Normalization}

With a value for the mean environmental density calculated for each
object in the Optical-9hr field it is possible to examine the 3D
density distribution across the entire redshift range. Due to the
flux-density limit of the observations and the much larger volume
being sampled at higher redshift, a peak in the density distribution
is found at $z \sim 0.4$ corresponding to the peak in the galaxy
number density. This peak in the detection rate would naturally lead
to an increase in the mean density being returned by the VT algorithm
for redshift slices in this range. Therefore two Voronoi cells from
two different redshift slices cannot be accurately compared in terms
of their environment. In order to counteract this bias it is necessary
to normalize the Voronoi cell areas across the entire redshift range
to produce a normalized environmental density for each object.

We therefore create a separate random field by applying a random
position to each Optical-9hr object within the H-ATLAS SDP
region within each redshift slice of that MC realisation.  We apply
our VT algorithm to the random field and determine the mean cell area
$\bar{x}_{Slice}$ (defined as the sum of the individual cell areas in
that slice divided by their number) and the standard deviation
$\sigma$ of each {\em{random}} redshift slice. Our density measure is
therefore given by,

\begin{equation}
S_{c}=\frac{\bar{x}_{slice}-x_{i}}{\sigma} \; ,
\label{eq:Sig} \end{equation}
where $x_{i}$ is the measured VT cell area for each object from the
real field per MC realisation and $S_c$ is the normalized density
value in comparison to a random distribution for each object.

$S_{c}$ therefore shows how the real field compares to a completely
random distribution, in terms of the standard deviation $\sigma$ of
that random distribution, and thus accounts for differences in
uniformity of the field between slices. This normalisation also allows
for the comparison of different objects from across redshift slices
with different population densities. Taking the mean of these values
across all MC realisations gives the mean normalized comparison
density ($\bar{S}_{c}$).

\subsection{Nearest-neighbour density comparison}\label{sec:NN}

\begin{figure}
\centering
\includegraphics[width=1.0\columnwidth]{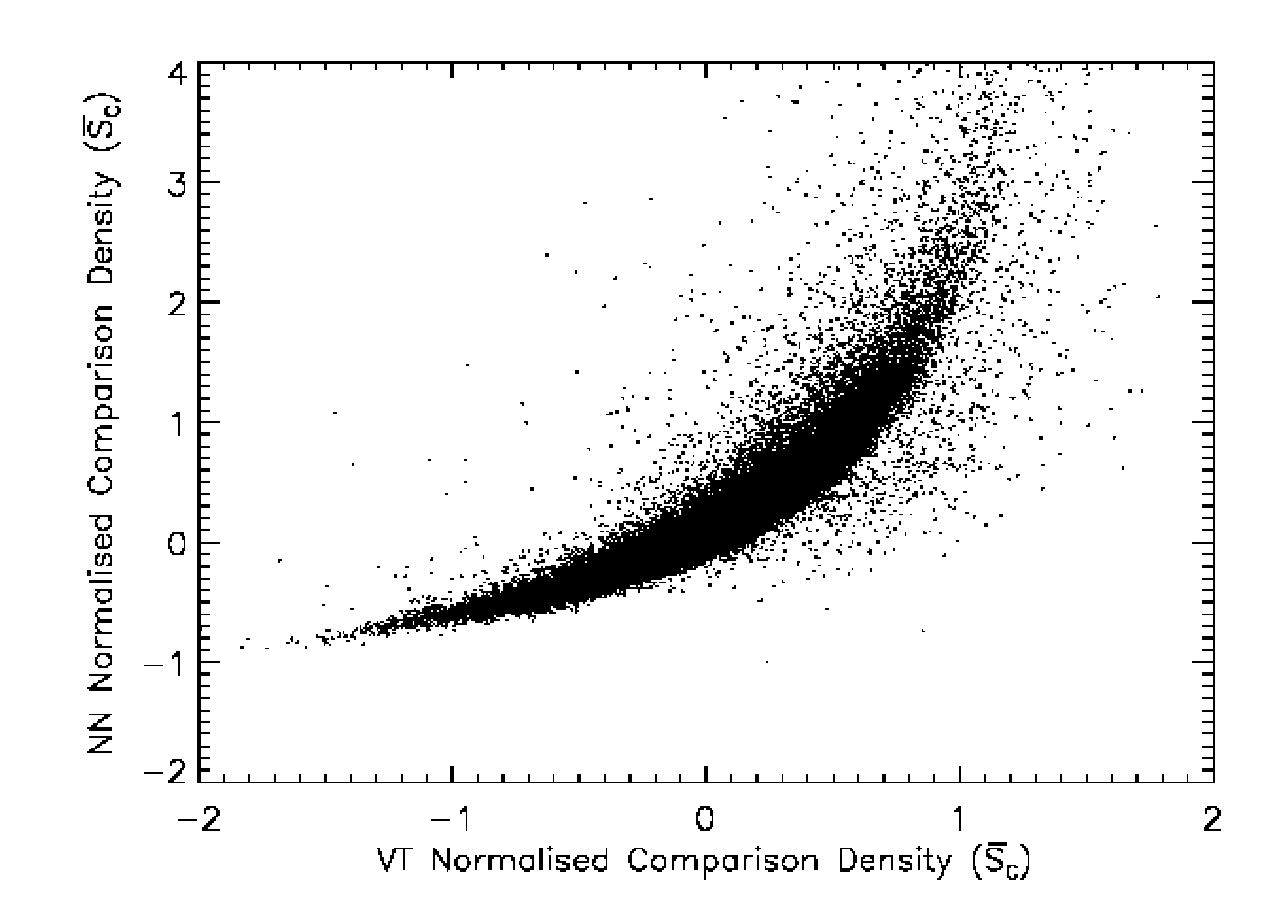}
\caption{The relationship between the normalized densities returned by
  the VT and NN techniques. The Spearman's Rank Correlation
  Coefficient ($r_{s}$) between both density distributions returns a
  value of $r_{s}=0.961$ at a $> 5\sigma$ level, indicating a strong
  correlation.}
\label{fig:VT_NN_correlation}
\end{figure}

Using VT to probe galaxy environments on individual galaxy scales is a
relatively new approach to the study of galaxy environmental density.
Previous work in the analysis of galactic environments has instead
predominantly relied on estimating the local density of galaxies using
the projected $N$th-nearest neighbour technique ($\Sigma_{N}$), which
measures the environmental density in terms of the number of galaxies
within a circular region defined by the radius to the $N$th-nearest
galaxy (e.g., \citealt{Dressler1980}; \citealt{Lewis_et_al2002};
\citealt{Miller_et_al2003}; \citealt{Balogh_et_al2004};
\citealt{Cooper_et_al2005}; \citealt{Silverman_et_al2009};
\citealt{Cucciati_et_al2010}; \citealt{Hernandez-Fernandez_et_al2012};
\citealt{Wijesinghe_et_al2012}). We therefore test our Voronoi
Tessellation density measure (VT) against this $N$th-nearest neighbour
technique (NN) in order to establish whether there are any significant
differences between the results obtained from both techniques, both in
terms of how our overall density correlations are affected as well as
a comparison of the techniques ability to probe detailed structure.

For our comparison we use $N = 5$ in line with the majority of recent
studies that have used the NN technique to examine local environmental
densities of galaxies (e.g., \citealt{Cucciati_et_al2010};
\citealt{Wijesinghe_et_al2012} and
\citealt{Hernandez-Fernandez_et_al2012}). We use the NN algorithm in
exactly the same way as our VT method described in Section
\ref{sec:Voronoi Tesselation Code}, with the NN algorithm applied to
each redshift slice within each MC cube, once more normalising the
field to account for differences in number density and uniformity
across the redshift range.  The only difference between the methods
comes as a result of the fact that, unlike a VT cell, the $\Sigma_{N}$
parameter represents larger densities with larger values and thus, to
reflect this, we reverse the sign of $\bar{S}_{c}$ such that positive
values once more represent a normalized overdensity.  We maintain the
boundaries applied to our VT density calculation to both prevent such
edge effects and to maintain comparison accuracy between the two
methods.

\begin{figure}
\centering
\includegraphics[width=1.0\columnwidth]{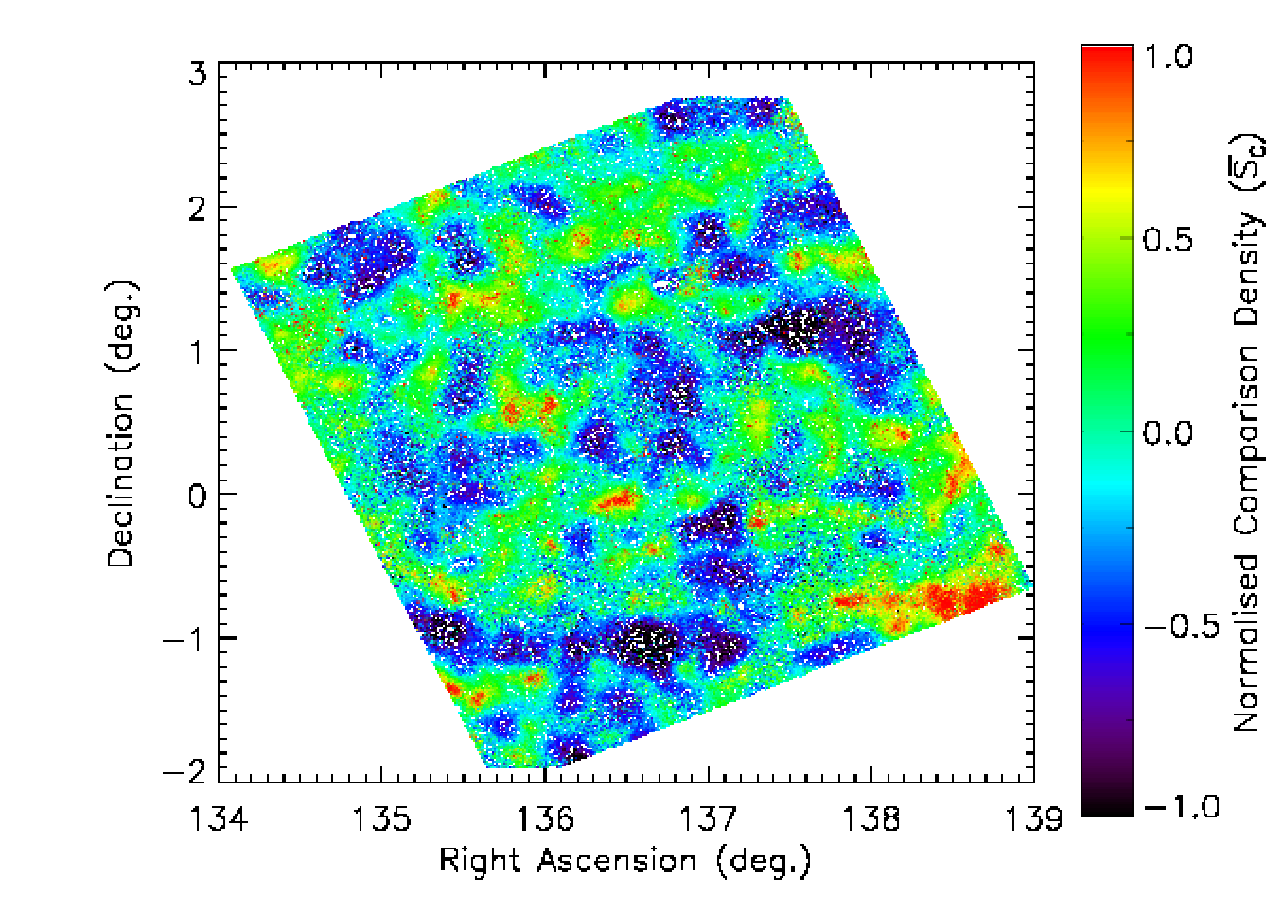}
\includegraphics[width=1.0\columnwidth]{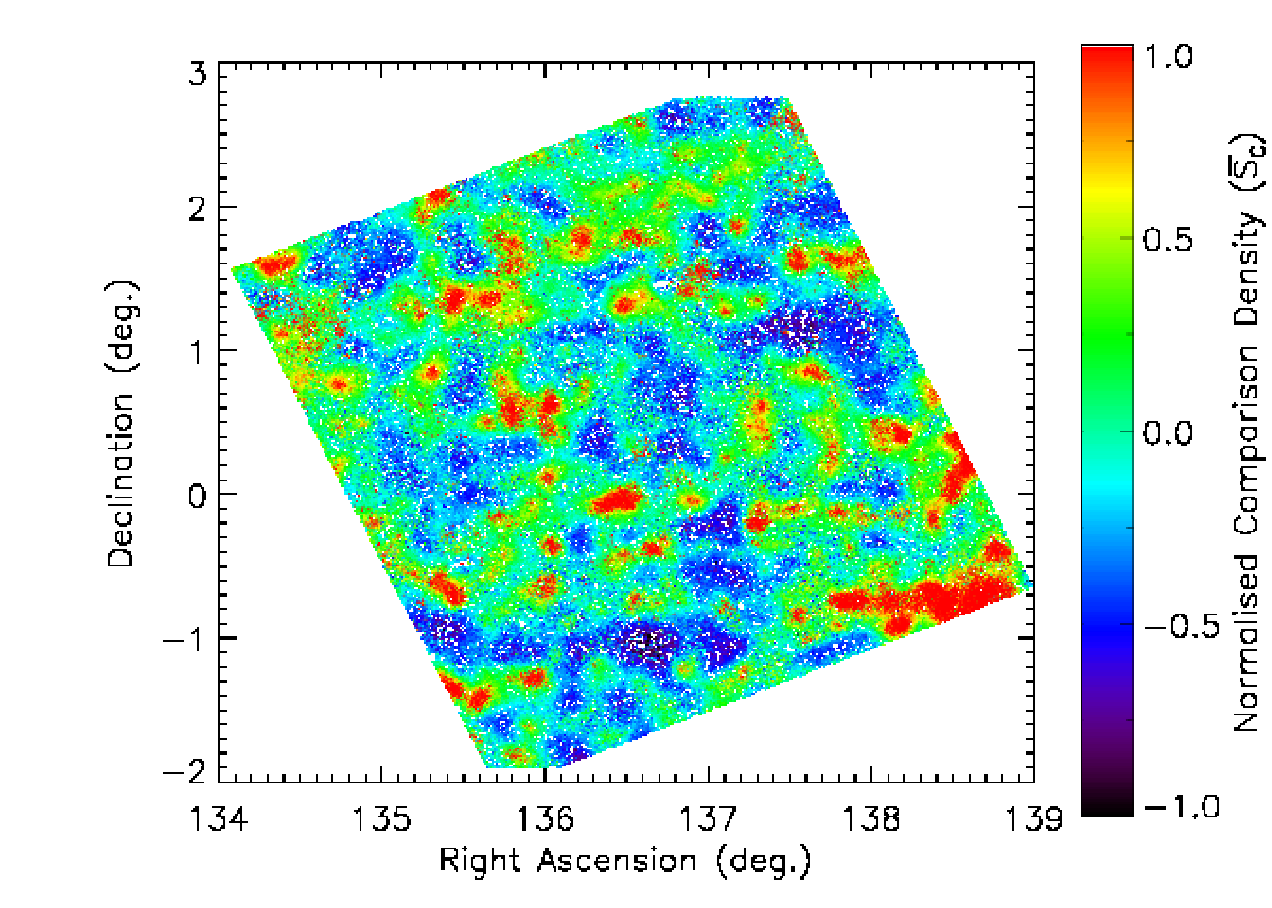}
\caption{Colour co-ordinated plots of the field with all redshift
  slices compiled to show the whole density distribution. Red and
  orange colours represent the most overdense regions (positive
  $\bar{S}_{c}$ values) and blue and purple represents the most
  underdense regions (negative $\bar{S}_{c}$ values) with the range of
  the normalized density limited to $-1 \leq \bar{S}_{c} \leq 1$ for
  clarity. {\em{Top}}: The VT method density output. {\em{Bottom}}:
  5th-nearest neighbour method density output. The plots confirm that
  the VT and NN methods reproduce the same density structure across
  the field.}
\label{fig:Realfield_NN}
\end{figure}

We apply both the Voronoi Tessellation and $N$th-nearest neighbour
techniques independently to our Optical-9hr catalogue;
Figure~\ref{fig:VT_NN_correlation} shows the relationship between the
initial outputs returned by both techniques. Calculating the
Spearman's Rank Correlation Coefficient ($r_{s}$) between both density
distributions returns a value of $r_{s}=0.961$ at a $> 5 \sigma$
level, indicating a strong (although non-linear) correlation. In
addition, Figure~\ref{fig:Realfield_NN} shows the normalized
comparison densities ($\bar{S}_{c}$) returned by each technique
plotted according to RA and Dec positions and coloured according to
density. Both figures clearly show that each density measure has
successfully reproduced the same general density structure across the
field, with the most extreme over- and under-densities located by both
methods. However, there are some noticeable differences between the
methods indicating that the intensity of the local density in specific
regions differs between each method.


From the direct comparison between the two methods in
Figure~\ref{fig:VT_NN_correlation}, it is clear that the NN method has
a greater dynamic range in the densest environments where the VT
method saturates. Conversely the VT method appears more suited to
distinguishing between less dense environments where the NN method
saturates. Figure~\ref{fig:Realfield_NN} shows that using the NN
method results in larger regions of peak overdensities with less
defined regions of intermediate density in comparison with the
structure distribution from the VT method. A full investigation of the
pros and cons of different density measures has been conducted by
\citet{Muldrew_et_al2012} and we refer the reader to that paper for
more information. But to summarise they find that the NN technique is
very poorly correlated with the respective dark-matter-halo mass,
although the NN technique is able to describe the internal densities
of high-mass haloes.


It is also clear that the initial value selected for $N$ will
determine the accuracy of the NN technique in various
environments. Where the value of $N$ remains below the number of
associated satellites, the measured density will increase with
increasing values of $N$. Subsequently in our comparison with the VT
method, the peak over densities returned from the $5$th-nearest
neighbour would be reduced if, for example, only the $3$rd-nearest
neighbour were used. However, with a larger value of $N$ the NN method
will lose resolution and become more susceptible to the projected
separations between distinct overdense regions, influencing the
density result.

In contrast, the VT method does not suffer from these issues, as
essentially the number of neighbours used to define the density are
not fixed. From the methodology of using the VT algorithm (Section
\ref{sec:Voronoi Tesselation Code}) it is evident that one does not
need to necessarily categorise each galaxy into a group or cluster,
but can instead simply measure the surface density of that galaxy
directly from the properties of its Voronoi cell. Consequently the VT
method is fully adaptable to changes in uniformity of the field and
calculates densities on individual galaxy scales. Therefore the VT
method represents a reliable and accurate alternative to the more well
established NN density measure.


\section{ANALYSIS}\label{sec:Analysis} 

\subsection{Far-infrared and control samples}\label{sec:KS-testing0}

As described in Section~\ref{sec:IR Data}, we use the likelihood-ratio
technique of \citet{Smith_et_al2011} to associate the far-infrared
sources with their optical counterparts.  This cross-matched sample is
hereafter named `FIR' (consisting of 2,265 objects) while
simultaneously removing them from the Optical-9hr catalogue reducing
this sample to $127,250$ objects (hereafter named `Optical'). These
sub-samples are shown in Table \ref{tab:Optical-9hr_Breakdown}, where
they are also divided according to the number of galaxies with
photometric and spectroscopic redshifts.

\begin{table}
  \caption{\label{tab:Optical-9hr_Breakdown} The number of objects within the Optical and FIR sub-samples of the initial $129,515$ objects of the Optical-9hr catalogue. These sub-samples are also divided according to the number of objects with photometric or spectroscopic redshifts in the density analysis.}
\begin{tabular*}{8.44cm}{lcccc}
  \hline
 Sample & Total Number & Number of Photo-z  & Number of Spec-z &\\
  \hline
  FIR & 2,265 &1,489 & 776 &\\ 
  Optical & 127,250 & 123,730 & 3,520 &\\ 
\hline
\end{tabular*}
\end{table}

In order to accurately compare how $\bar{S}_{c}$ values differ between
the FIR and Optical catalogues it is necessary to ensure that we are
comparing like with like, such that the objects selected for
comparison should be considered to be from the same population. By
selecting a matched sample of galaxies based on their colour, SDSS
$r-$band model apparent magnitude and redshift distributions
we ensure that these properties have no influence on any differences
in environmental density found between the two catalogues.
This is achieved by gridding the field in four dimensions in order to
incorporate all $g-r$, $r-i$, $m_{r}$ and $z$ parameter space,
selecting matched objects as only those which share an associated grid
space in all four planes. The choice of $g-r$ and $r-i$ colours are
selected as we limit our Optical catalogue in the $r$-band apparent
magnitude. As explained in Section \ref{sec:Optical Data} and as shown
in \cite{Taylor2011}, colours provide a reasonable method of matching
sources in terms of their stellar mass over the redshift range under
investigation here, although we also investigate this further in
Section~\ref{sec:dustreddening}. The grid elements applied to the
total $g-r$ and $r-i$ colour distributions incorporate $0.1$ and
$0.06$ magnitudes respectively in colour space. This difference
reflects the larger range of the total $g-r$ colour
distribution. Simultaneously, the $z$ and $m_{r}$ ranges have grid
elements incorporating $0.02$ in redshift and $0.38$ in $r-$band
magnitude.

All Optical sources that share an associated grid space with a FIR
detection in all four planes are initially grouped as potential
matches to those FIR objects. Then, within each grid space, a multiple
of the potential matches (totalling three times the FIR sources in
that grid space) are selected as matched objects. Selecting Optical
matches equal to a multiple of the FIR sources in each grid element
allows for a more robust comparison without sacrificing any
similarities between their distributions. Any additional Optical or
FIR sources that are not matched are discarded. The FIR sample
contains considerably fewer objects than the Optical sample ($2,265$
against $127,250$), therefore a large proportion of the Optical sample
will not have an associated FIR object and thus will be lost from the
final cross-match. This reduces the sample size to $2,706$ and $902$
for the Optical and FIR samples respectively.

In addition, the number-density of galaxies reduces with increasing
redshift to such an extent as to affect our sampling. Therefore we
apply a maximum peak redshift limit onto the samples of $z \leq 0.5$.
This maximum redshift does not influence the $\bar{S}_{c}$ values of
the remaining sample due to the fact that each $\bar{S}_{c}$ value
already incorporates the high redshift galaxies via the full z-PDF
sampling achieved within the algorithm. Figures~\ref{fig:GRvsRI}
\&~\ref{fig:ZvsMag_Rmatch} show the colour, redshift and magnitude
distributions for these two matched samples.


\subsection{Comparison of the FIR and Optical samples}\label{sec:KS-testing}

\begin{figure}
\centering
\includegraphics[width=1.0\columnwidth]{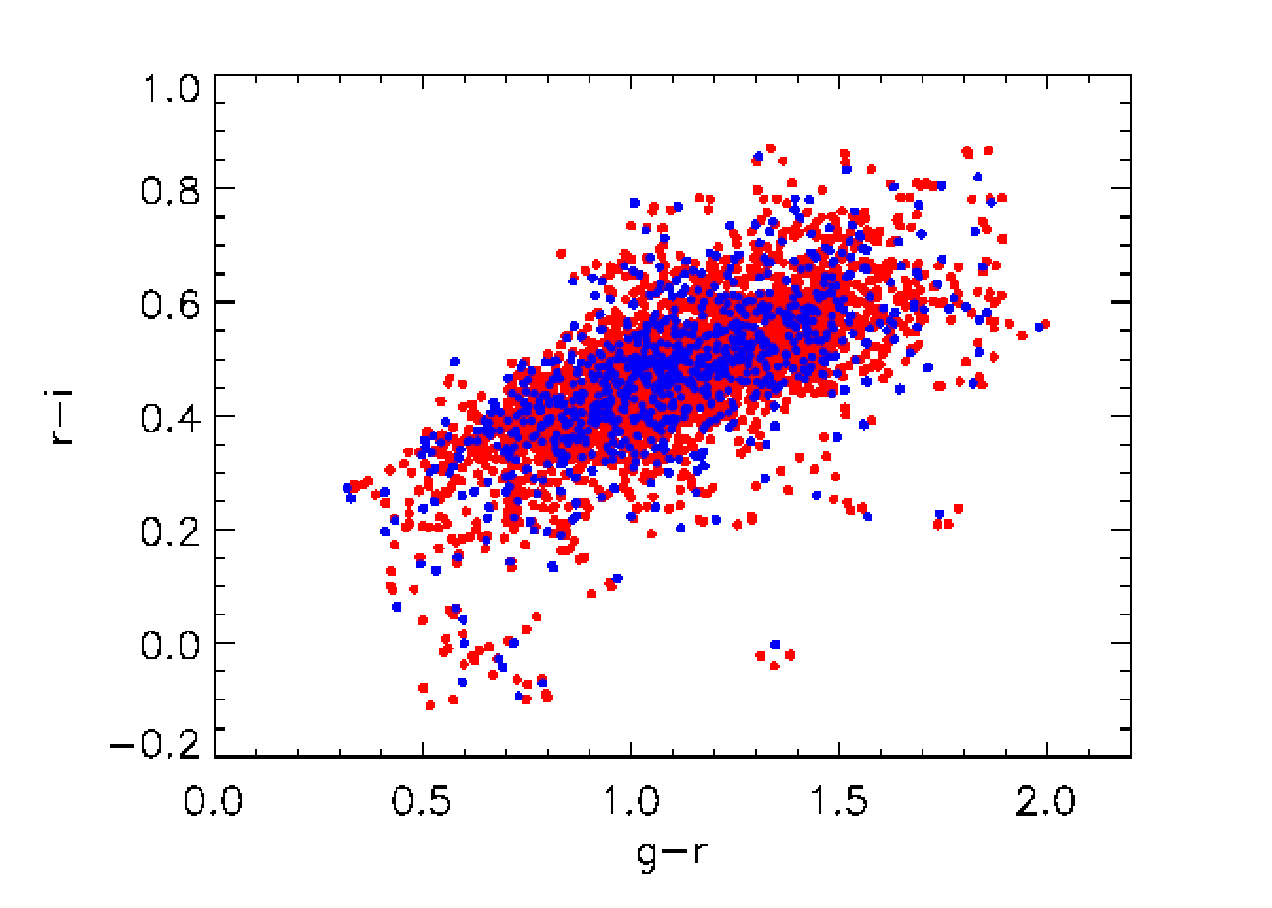}
\caption{g-r vs r-i colour distribution for the `matched'
  catalogues Optical ({\em red}) and Herschel
  ({\em{blue}}) numbering $2,706$ and $902$ sources respectively.}
\label{fig:GRvsRI}
\end{figure}

\begin{figure}
\centering
\includegraphics[width=1.0\columnwidth]{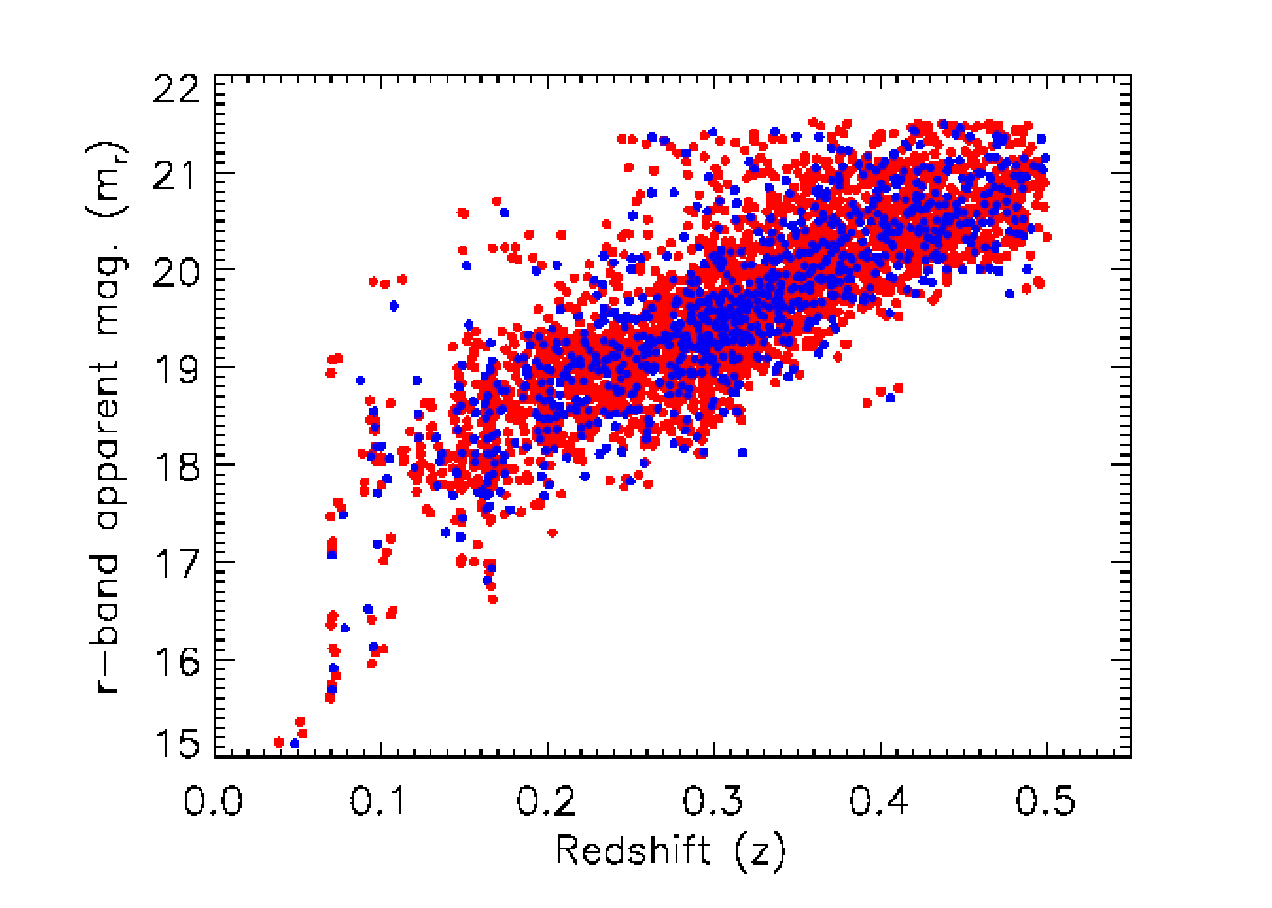}
\caption{Redshift vs $r-$band apparent magnitude ($m_{r}$) for the
  `matched' Optical ({\em red}) and Herschel ({\em{blue}})
  catalogues. Only the redshift range of $0 < z \leq 0.5$ and $m_{r}$
  range of $15<m_{r}<21.5$ was included in the sampling, outside of
  these ranges the completeness of the catalogues reduced
  significantly.}
\label{fig:ZvsMag_Rmatch}
\end{figure}


We use one- and two-dimensional Kolmogorov-Smirnoff tests (KS-tests)
to confirm that our matched samples are consistent with having been
drawn from the same underlying distribution in terms of their colour,
magnitude and redshift distributions and that the various combinations
are consistent with each other.
These tests demonstrate that our null hypothesis, such that the
`Optical' control sample is drawn from the same underlying
distribution as the FIR sample, cannot be rejected at a significant
level.
The results of these tests are presented in Table~\ref{tab:Results}.
We then applied the one-dimensional KS test to the environmental
density measurements for the Optical and FIR samples ($\bar{S}_{c}$),
which return a probability of just $4.2 \times 10^{-4}$, rejecting the
null hypothesis of them being drawn from the same underlying
distribution at the $3.5\sigma$ level.
Therefore we find a significant difference in the distribution of the
galaxy environmental density between far-infrared selected galaxies
and a control sample with no detectable far-infrared emission.
To test this result further we applied a two-dimensional KS-test to both
Optical and FIR populations comparing the environmental densities in
conjunction with the optical properties and the redshifts. The
two-dimensional KS-test comparing the colours, $r-$band magnitude and
redshift distributions to the environmental density values are
presented in Table~\ref{tab:Results} and show that the environmental
densities of the FIR and Optical samples are significantly different
in all cases.

\begin{table}
\raggedright
\caption{\label{tab:Results} Two sample and two-dimensional KS and MWU-test results over the full SFR and redshift range ($0 < z \leq 0.5$). Where {\em{op}} represents Optical ($2,706$ objects) and {\em{FIR}} represents FIR ($902$ objects). The two density distributions are significantly different at the 3.5$\sigma$ level from KS tests, with the medians of the distributions different at the 4.5$\sigma$ level from MWU tests.}
\begin{tabular*}{8.44cm}{lcc}
  \hline
  Distributions Compared & KS Prob. & MWU Prob.\\
  \hline
  $z_{op}$ vs $z_{FIR}$ & 0.999 & 0.403 \\
  $(g-r)_{op}$ vs $(g-r)_{FIR}$ & 0.974 & 0.374 \\
  $(r-i)_{op}$ vs $(r-i)_{FIR}$ & 0.395 & 0.239 \\
  $m_{r(op)}$ vs $m_{r(FIR)}$ & 0.719 & 0.290 \\
  $(\bar{S}_{c})_{op}$ vs $(\bar{S}_{c})_{FIR}$ & $<10^{-3}$ & $<10^{-5}$ \\
  $(g-r, r-i)_{op}$ vs $(g-r, r-i)_{FIR}$ & 0.224 & - \\
  $(g-r, z)_{op}$ vs $(g-r, z)_{FIR}$ & 0.956 & -\\
  $(r-i, z)_{op}$ vs $(r-i, z)_{FIR}$ & 0.468 & -\\
  $(m_{r}, z)_{op}$ vs $(m_{r}, z)_{FIR}$ & 0.755 & - \\
  $(g-r, m_{r})_{op}$ vs $(g-r, m_{r})_{FIR}$ & 0.839 & -  \\
  $(r-i, m_{r})_{op}$ vs $(r-i, m_{r})_{FIR}$ & 0.339 & -\\
  $(g-r, \bar{S}_{c})_{op}$ vs $(g-r, \bar{S}_{c})_{FIR}$ & 0.005 & - \\
  $(r-i, \bar{S}_{c})_{op}$ vs $(r-i, \bar{S}_{c})_{FIR}$ & 0.005 & - \\
  $(m_{r}, \bar{S}_{c})_{op}$ vs $(m_{r}, \bar{S}_{c})_{FIR}$ & 0.006 & - \\
  $(z, \bar{S}_{c})_{op}$ vs $(z, \bar{S}_{c})_{FIR}$ & 0.010 & - \\  
  \hline
\end{tabular*}
\end{table}

\begin{table}
  \caption{\label{tab:Results2} Full SFR range KS-test and MWU-test
    results for the comparison of both the Optical and FIR populations
    $\bar{S}_{c}$ distributions within individual redshift
    slices. From KS results the density distributions are different
    between the 2.2$\sigma$-3.3$\sigma$ level. The number of objects
    from each population are given:}
\begin{tabular*}{8.698cm}{lcccc}
  \hline
  Redshift Slice & Optical & IR & KS Prob. & MWU Prob.\\
  \hline
  $0 \leq z < 0.25$ & 680 & 227 & 0.028 & 0.028 \\ 
  $0.25 \leq z < 0.50$ & 2,026 & 675 & $<10^{-3}$ & $<10^{-4}$ \\ 
\hline
\end{tabular*}
\end{table}

\begin{figure*}
 \begin{center}
\includegraphics[width=0.693\columnwidth]{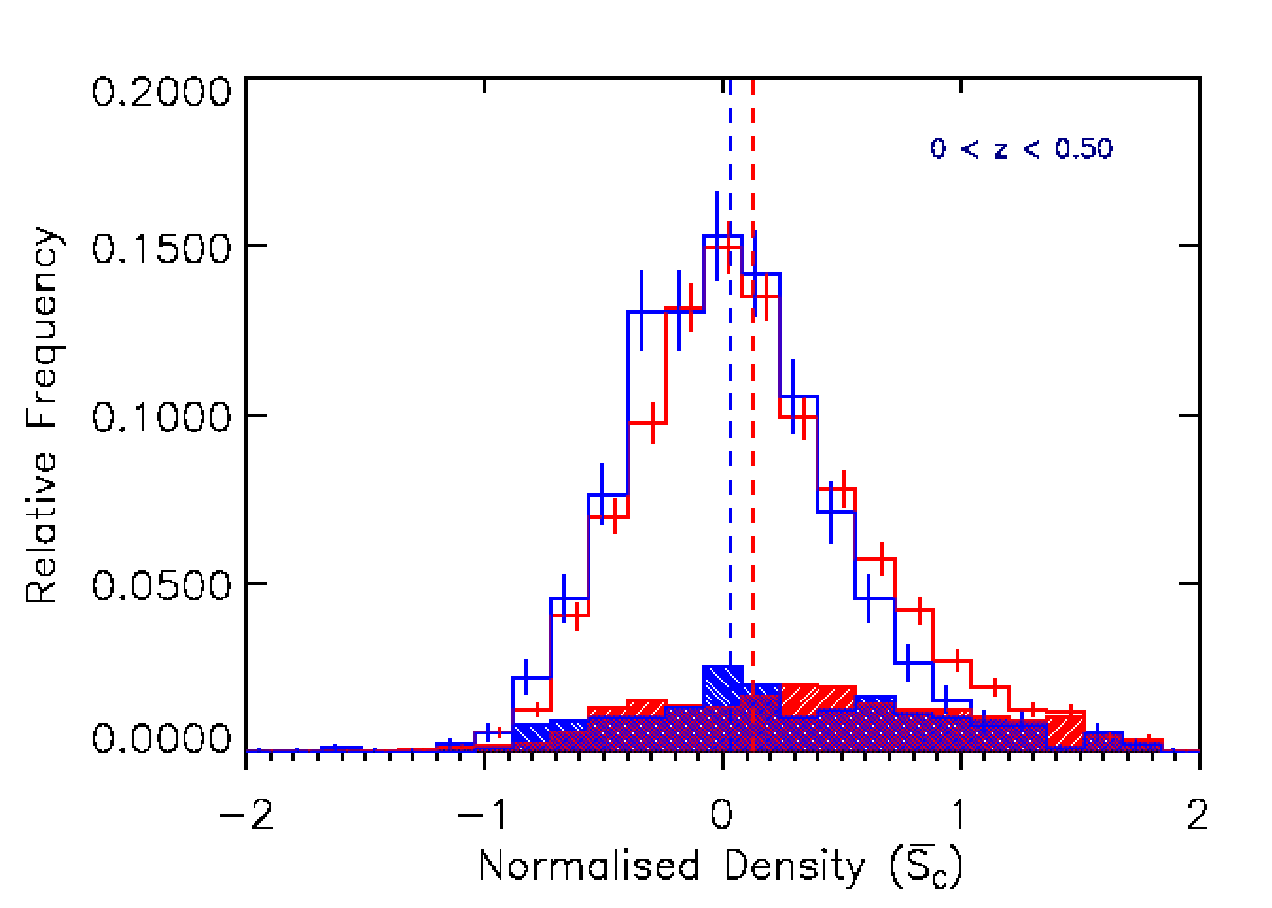}
\includegraphics[width=0.693\columnwidth]{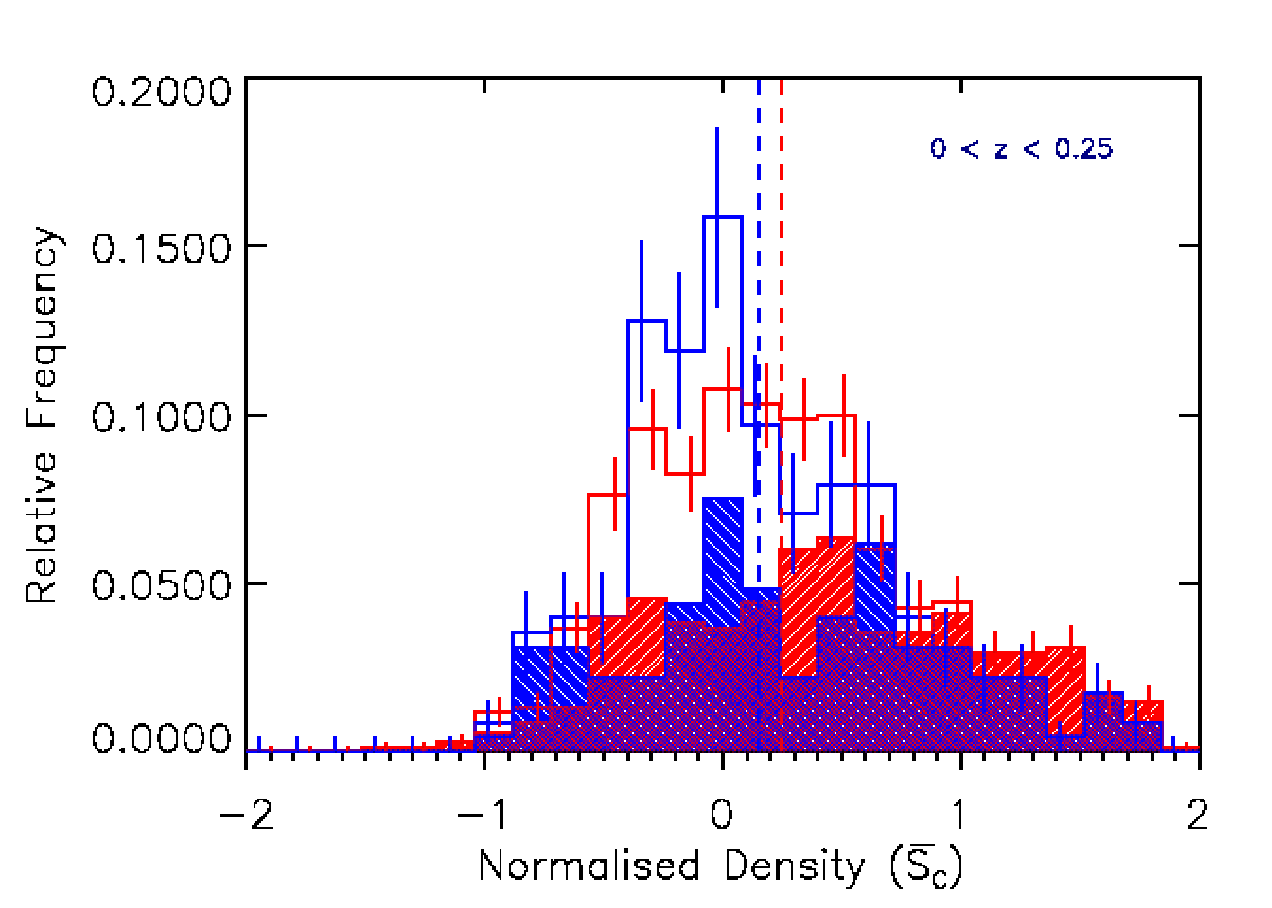}
\includegraphics[width=0.693\columnwidth]{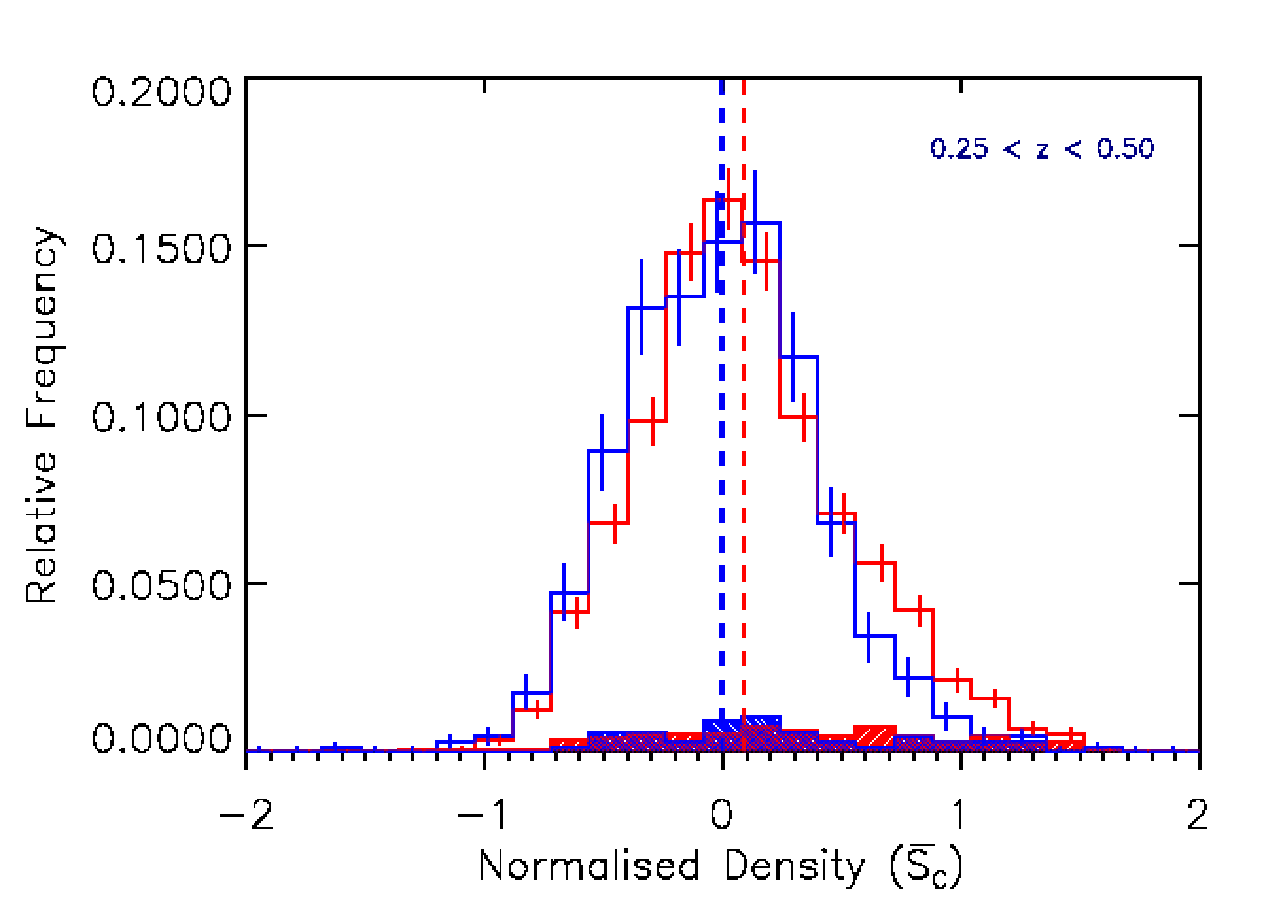}
\end{center}
\caption{Normalized histograms showing how the distribution of
  environmental density ($\bar{S}_{c}$) of the total colour matched
  Optical ({\em{red}}) and FIR ({\em{blue}}) populations compare,
  with shaded histograms representing the number of spectroscopic
      redshifts in each sample. {\em{Left}}: Full matched catalogues
  containing $2,706$ Optical and $902$ FIR objects. The histograms
  show error bars depicting the normalized error associated with each
  bin, where $\bar{S}_{c}>0$ signifies an overdensity and
  $\bar{S}_{c}<0$ signifies an underdensity in terms of the entire
  redshift range ($0 \leq z < 0.5$). The FIR data are shifted generally
  to lower $\bar{S}_{c}$ values, with the mean of the distribution at
  ($3.46 \pm 1.49$) $\times 10^{-2}$ ({\em{blue dashed line}}), this
  is contrasted against the mean of the Optical distribution at
  ($12.70 \pm 0.95$) $\times 10^{-2}$ ({\em{red dashed line}}). KS and
  MWU-tests indicate a significant difference to the 3.5$\sigma$
  level.
  {\em{Centre}} and {\em{Right}}: Normalized histograms that show the
  full matched sample binned in redshift ($0 \leq z < 0.25$) and
  ($0.25 \leq z < 0.50$) respectively, showing a continued separation
  between the distributions increasing with redshift from 2.2$\sigma$
  to 3.3$\sigma$ significance.}
 \label{fig:Ho4norm}\label{fig:Hofullnorm} 
 \end{figure*}

 The extent of this difference is illustrated in
 Figure~\ref{fig:Hofullnorm} where we show normalized histograms of
 the two environmental density distributions. It is clear that the
 Optical population ({\em{red}}), with the mean of its distribution at
 $\bar{S}_{c}=(12.70 \pm 0.95) \times 10^{-2}$ (denoted by the red
 dashed line), is more overdense (has larger values of $\bar{S}_{c}$)
 than the FIR population ({\em{blue}}), with the mean of its
 distribution at $\bar{S}_{c}=(3.46 \pm 1.49) \times 10^{-2}$ (denoted
 by the blue dashed line).  Using the Mann-Whitney U (MWU) test we
 also test for differences between the median values of the
 distributions. For the Optical and FIR populations the test returns a
 probability of $5 \times 10^{-6}$, indicating that the two
 populations have significantly different median values at the
 $4.5\sigma$ level.
In Section \ref{sec:NN_APP} we show the same result is returned when
the NN method is used to calculate the environmental density, thus
showing consistency with our VT result.

\subsection{Redshift binning}\label{sec:z-splitting}

Clearly in any flux-density limited sample there is a bias in the
sense that the higher-redshift sources are more luminous than those at
lower redshift. Therefore, to further examine the difference between
the Optical and FIR environmental density distributions we split the
two populations into two redshift slices of $0 < z \leq 0.25$ and
$0.25 < z \leq 0.5$.
The $0 \leq z \leq 0.25$ bin contains $680$ objects from the Optical
population with $227$ objects from the FIR population and the $0.25 <
z \leq 0.5$ bin contains $2,026$ objects from the Optical population
and $675$ objects from the FIR population. The results of the KS-tests
comparing the density measurements within these bins are shown in
Table~\ref{tab:Results2}; these show that the null hypothesis can be
rejected and the two populations can be considered different in terms
of their overall density distributions at the 2.2$\sigma$ level for
the low-redshift bin and $3.3\sigma$ level for the high-redshift
bin. The MWU test returns probabilities of $0.028$ and $3.4 \times
10^{-5}$ for the low- and high-redshift bins respectively, also
indicating that the two distributions have significantly different
median values. These binned distributions are shown in
Figure~\ref{fig:Hofullnorm}. Consistent results are found when the NN
method is used to calculate the environmental density, as shown in
Section \ref{sec:NN_APP}.
 
\begin{table*}
\raggedright
\caption{\label{tab:Results5SFR} Two sample KS and MWU-test results for each
  SFR bin, collectively over the full redshift range ($0 < z \leq 0.5$)
  where {\em{op}} represents the Optical and {\em{FIR}} represents
  the FIR populations. (A): SFR of 0 to 15 $\Msolar$\,yr$^{-1}$ contains
  $414$ cross-matched Optical objects and $138$ cross-matched FIR objects. %
  (B): SFR of 15 to 30 $\Msolar$\,yr$^{-1}$ bin contains
  $1,104$ cross-matched Optical objects and $368$ cross-matched FIR objects. %
  (C): Minimum SFR of 30 $\Msolar$\,yr$^{-1}$ bin containing $879$ cross-matched Optical objects and $293$
  cross-matched FIR objects. 
}
\begin{tabular*}{17.92cm}{lcccccc}
\hline
 Distributions Compared & KS Prob. (A) & MWU Prob. (A) & KS Prob. (B) & MWU Prob. (B) & KS Prob. (C) & MWU Prob. (C)\\
\hline
$z_{op}$ vs $z_{FIR}$ & 0.999 & 0.484 & 0.911 & 0.359 & 0.324 & 0.372\\
$(g-r)_{op}$ vs $(g-r)_{FIR}$ & 0.834 & 0.357 & 0.964 & 0.420 & 0.997 & 0.420\\
$(r-i)_{op}$ vs $(r-i)_{FIR}$ & 0.054 & 0.057 & 0.998 & 0.453 & 0.999 & 0.474\\
$m_{r(op)}$ vs $m_{r(FIR)}$ & 0.994 & 0.410 & 0.994 & 0.368 & 0.630 & 0.310\\
$(\bar{S}_{c})_{op}$ vs $(\bar{S}_{c})_{FIR}$ & 0.009 & 0.009 & 0.006 & $<10^{-3}$ & $<10^{-3}$ & $<10^{-5}$ \\
\hline
\end{tabular*}
\end{table*}

\begin{figure}
\centering
\includegraphics[width=1.0\columnwidth]{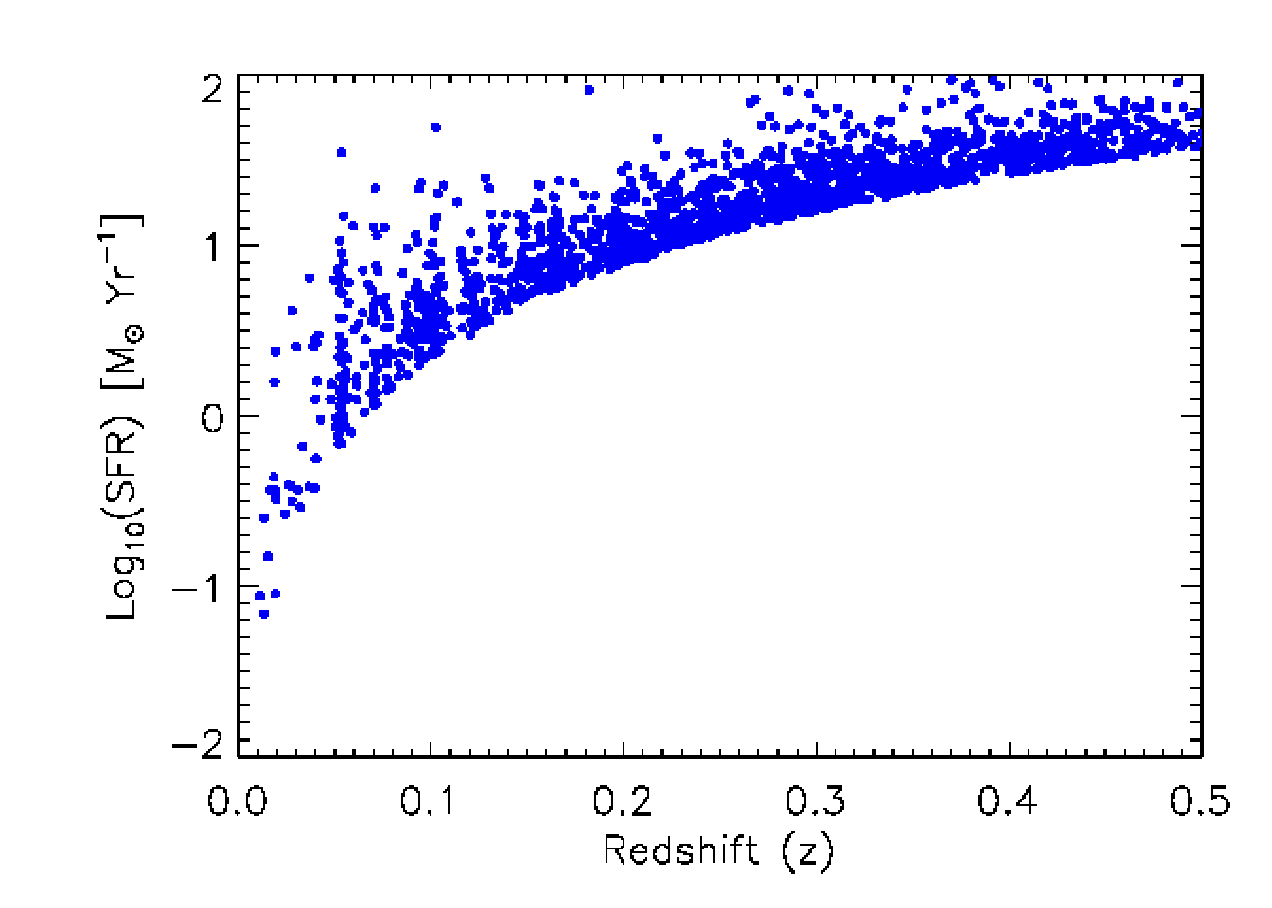}
\centering
\includegraphics[width=1.0\columnwidth]{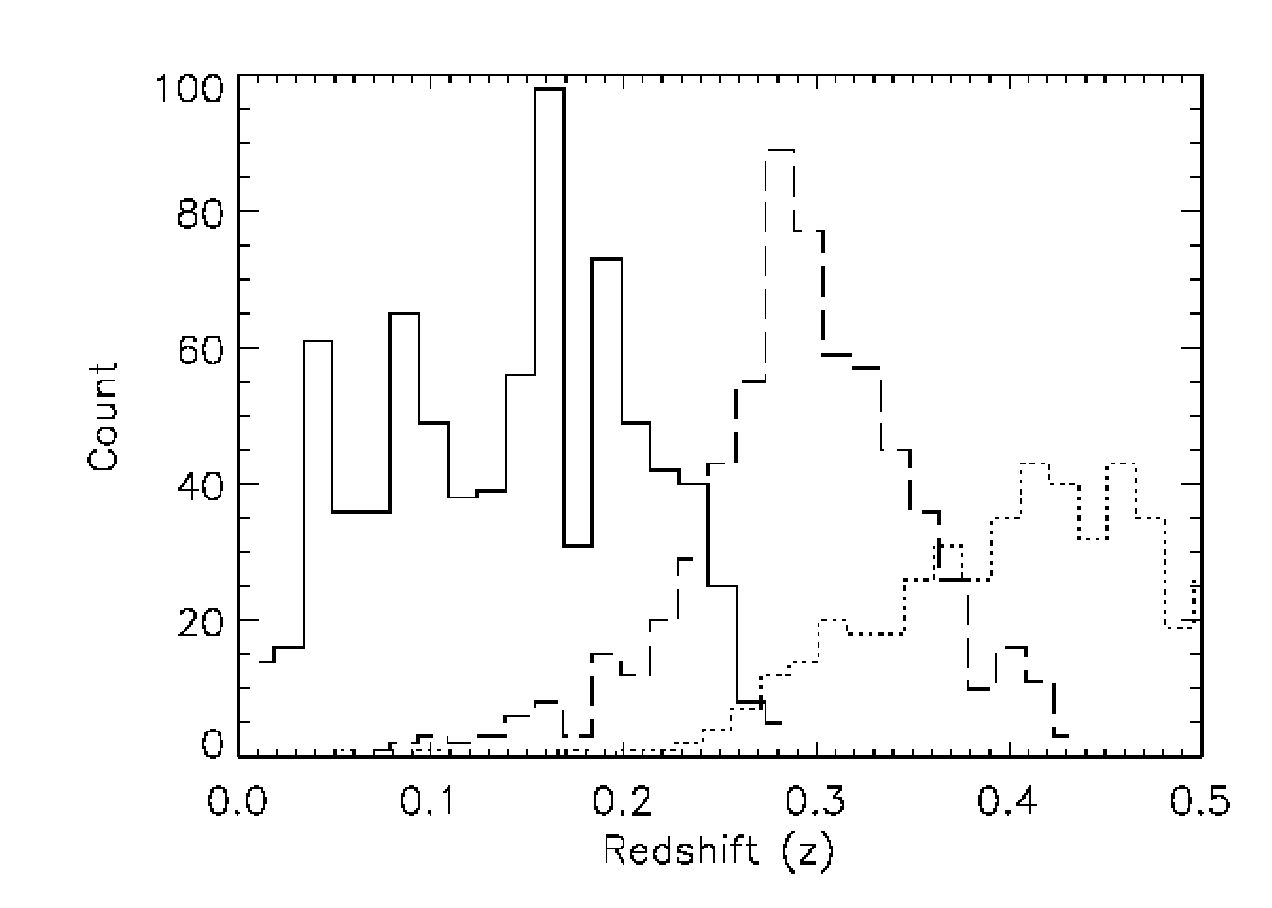}
\caption{{\em{Top}}: The calculated SFR ($\Msolar$\,yr$^{-1}$) against
  redshift for the total FIR catalogue. {\em{Bottom}}: The three SFR bins from the FIR
  catalogue versus redshift. The $0 < SFR < 15 \Msolar$\,yr$^{-1}$ bin (Solid
  line) containing $781$ objects, the $15 < SFR \leq 30 \Msolar$\,yr$^{-1}$ bin (Dashed line) containing $631$ objects and the $SFR >
  30 \Msolar$\,yr$^{-1}$ bin (Dotted line) containing $423$ objects
  $z\leq0.5$.}
\label{fig:ZvsSFR}
\end{figure}

These results show that, as with the full redshift range, objects in
both redshift bins are significantly different in terms of their
overall density distributions and their median values. However, it is
evident that these statistical differences are higher in the
higher-redshift bin and that this bin contains a larger number of
objects in both samples. We investigate the impact of this difference
in number density with increasing redshift by matching the number of
galaxies in the higher and lower bins and repeating the sample
comparisons. First, by increasing the redshift boundary between the
higher and lower bins (from $z = 0.25$ to $z = 0.32$) to achieve
approximate matching in sample sizes above and below this
redshift. Second, by reducing the number of objects within the higher
redshift bin to match exactly with the lower bin samples. In both
cases we find that the same trends are found and our results remain
the same.

With fewer objects in the lower redshift bin, the signal to noise will
be lower at these redshifts, flattening the density distributions and
affecting the comparison. Furthermore, at higher redshifts, objects
with lower IR luminosities are excluded by the flux-density limit of
the H-ATLAS survey. Thus, the density distribution of the less
luminous far-infrared galaxies may actually be similar to the Optical
population. In contrast the higher redshift bins contain a much higher
proportion galaxies with higher levels of star formation and therefore
exhibit a stronger correlation with density. The consequence of this
bias is that the statistical differences found between the total
Optical and FIR distributions (shown in Table \ref{tab:Results}) may
be being diluted by galaxies with low SFRs at low redshift. In order
to examine the full impact of these objects, in Section
\ref{sec:SFRlim}, we apply SFR bins to the FIR catalogue and repeat
the above analysis.

\subsection{Star-formation rate vs environmental density}\label{sec:SFRlim}

The SFR of a galaxy can be estimated using the relation given in
\citet{Kennicutt1998} as proportional to the total IR luminosity
($L_{FIR}$) over 8-1000$\micron$, assuming a Salpeter IMF between
$0.1\Msolar-100\Msolar$ \citep{Salpeter1955}, such that:

\begin{equation} SFR(\Msolar {\rm yr}^{-1})=4.5 \times 10^{-44} \cdot
  L_{\rm FIR} ({\rm ergs} \cdot {\rm s}^{-1}) \; ,  \label{eq:SFR} \end{equation} 
The thermal emission of far-IR galaxies can be represented by a
modified black body emission spectrum from \citet{Blain_et_al2003}:

\begin{equation}
f_\nu \propto \frac{\nu^{3+\beta}}{\exp\left(\frac{h\nu}{kT} -1\right )},
\end{equation}
where $h$ is the Planck constant, $k$ is the Boltzmann constant, and
$T$ represents the temperature. The emissivity index, $\beta$ modifies
the Planck function by assuming that the dust emissivity varies as a
power law with frequency, $\nu^{\beta}$, where $\beta$ can be between
1-2 as described in \citet{Hildebrand1983}, depending on the frequency
of the observations. We fix the dust emissivity index to $\beta = 1.5$
with dust temperature ($T$) equal to 26\,K as found by
\citet{Dye_et_al2010} and \citet{Jarvis_et_al2010} and integrate the
modified blackbody over the wavelength range $8-1000\mu$m to obtain
the far-infrared luminosity. We then use equation~\ref{eq:SFR} to
determine the SFR (in $\Msolar$~yr$^{-1}$) for each galaxy (see
Figure~\ref{fig:ZvsSFR}).

As a check of our $L_{FIR}$ values we compare the galaxies in our
sample to the subset of objects with far-infrared luminosities
determined from the full energy balance models of
\citet{dacunha_et_al2008} by \citet{Smith_et_al2012}. We find that our
$L_{FIR}$ values are slightly underestimated, and a correction factor
of $1.25$ is needed to produce a 1:1 correlation. This suggests that,
as we are assuming a fixed dust temperature of 26\,K, our estimate
misses $\sim25$ per cent of the total dust luminosity, and hence the
SFRs are underestimated in our calculation. We therefore apply this
correction factor to our $L_{FIR}$ values to account for this
difference in our resultant SFRs. We note that this correction makes
very little difference to our overall results on the relative
environmental densities between different bins in SFR.



It is also worth noting that the SFRsestimated from FIR emission may
be overestimates of the true SFR. We know that far-infrared emission
is a tracer of star formation in an idealised case where young stars
dominate the radiation field and dust opacity is high
(\citealt{Kennicutt_et_al2009}). Multi-temperature dust distributions
and emission from dust heated by older ISM stars
\citep{dacunha_et_al2008, Smith_et_al2012} are not be expected to be
consistent with equation \ref{eq:SFR}}, and this also plays a part in
our correction factor of 1.25. This must also be balanced against the
fact that any unobscured component of star formation would not be
detected in the FIR. Thus, although far-infrared emission is highly
correlated with SFR we note that the absolute values of SFR should be
used with caution.

\begin{figure}
\centering
\includegraphics[width=1.0\columnwidth]{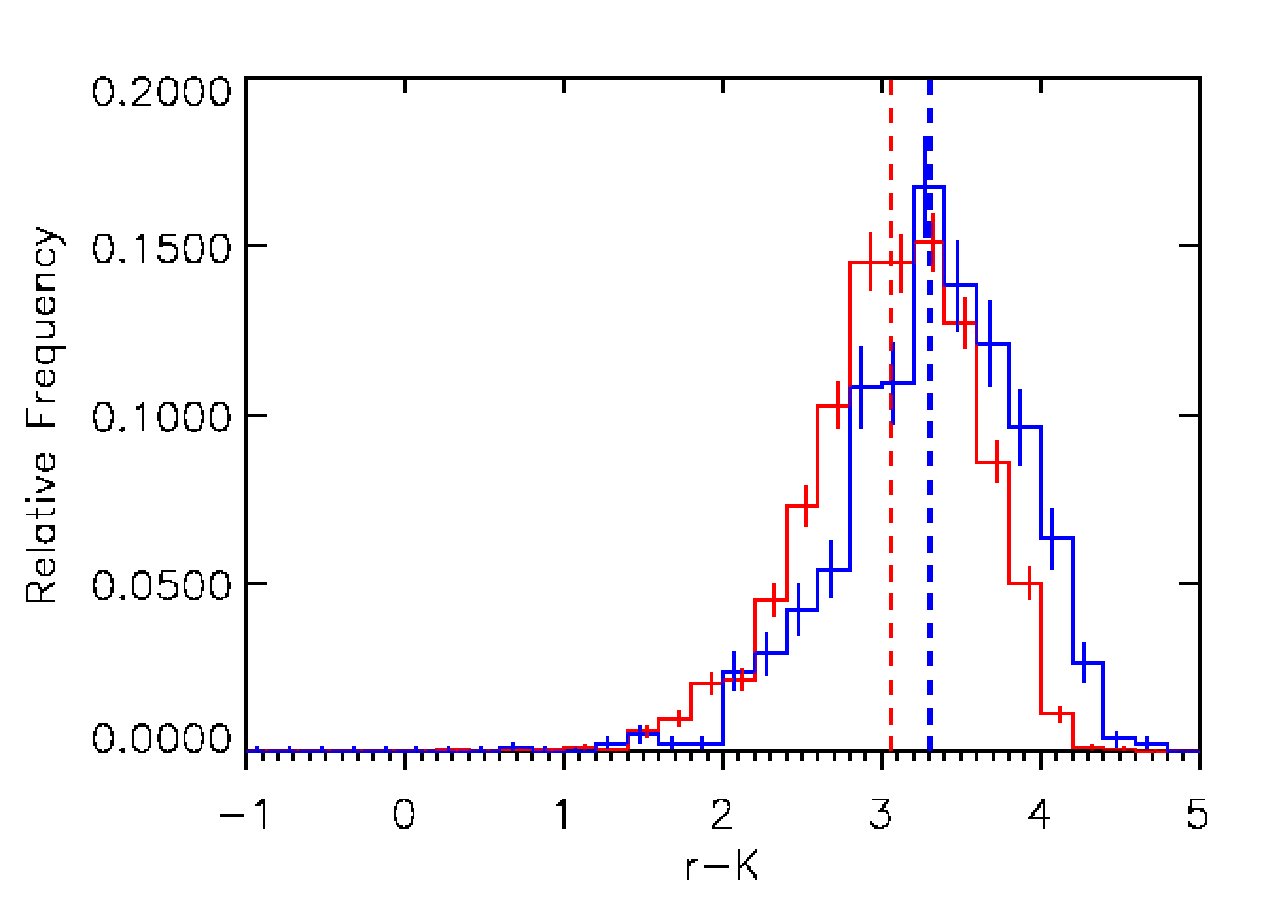}
\caption{Histograms of the $r-K$ colours for the Optical (red) and FIR
  (blue) samples showing that they have significantly different
  distributions at a $> 5\sigma$ level with the FIR galaxies lying
  redward of the Optical galaxies.}
\label{fig:rminK}
\end{figure}

\begin{figure}
\centering
\includegraphics[width=1.0\columnwidth]{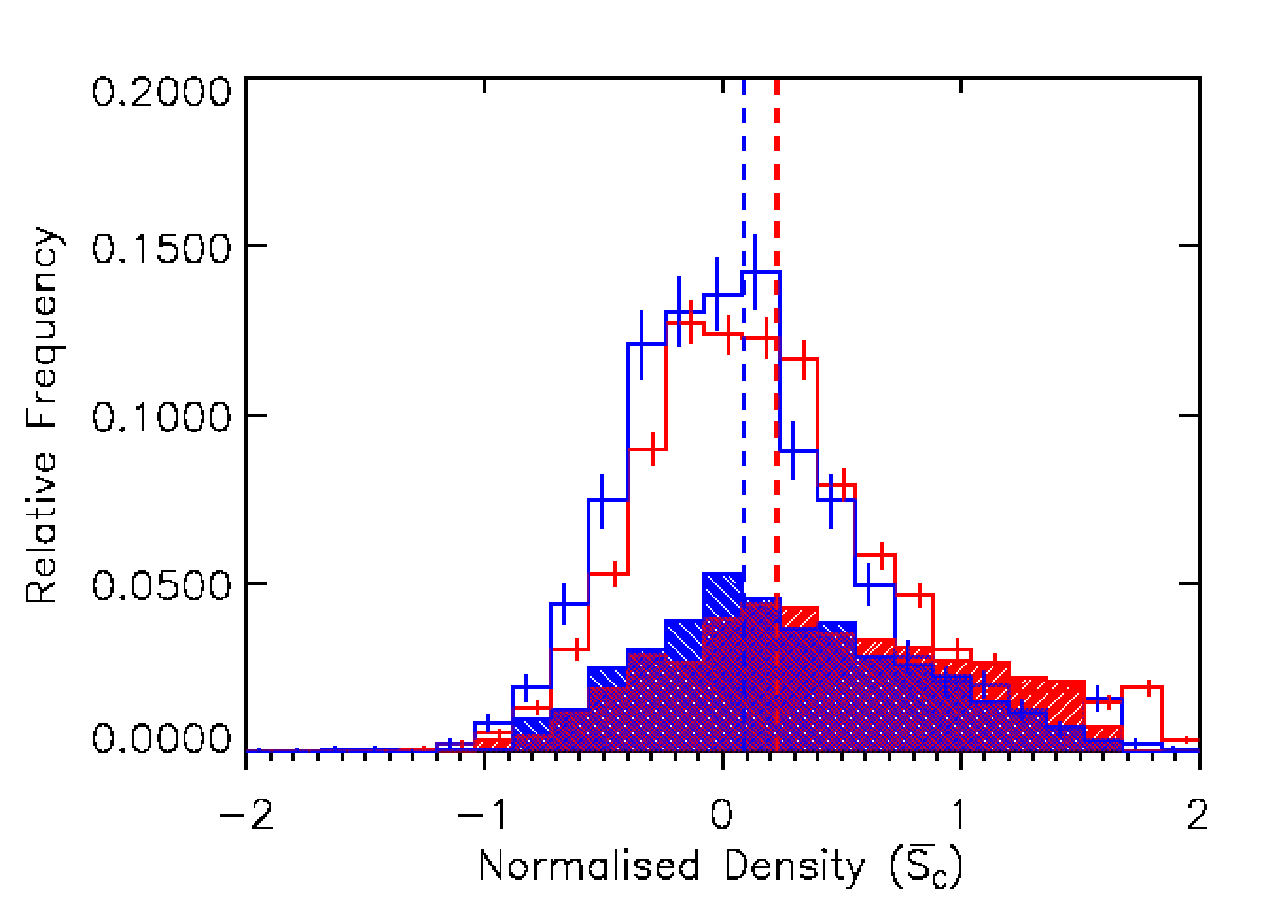}
\caption{Normalised histograms of the environmental density
  ($\bar{S_{c}}$) of both Optical ({\em{red}}) and FIR ({\em{blue}})
  populations, cross-matched in $r-K$, $m_{K}$ and $z$ parameter
  space. KS and MWU tests indicate a significant difference to the $>
  5 \sigma$ level. The FIR data is shifted generally to lower
  $\bar{S}_{c}$ values, with the mean of its distribution at $0.09 \pm
  0.02$ ({\em{blue dashed line}}), contrasted against the mean of the
  Optical distribution at $0.23 \pm 0.01$ ({\em{red dashed line}})
  Shaded histograms represent the number of spectroscopic redshifts in
  each sample.}
\label{fig:K2_histo}
\end{figure}

We bin the FIR objects in terms of their SFR in bins of $0-15$,
$15-30$ and $>30\Msolar$\,yr$^{-1}$ in order to compare the impact of
different SFRs on our initial results. Again matching the control
sample to the individual binned SFR samples we perform KS-tests and
MWU-tests on all combinations. The results of these tests are shown in
Table~\ref{tab:Results5SFR}. From KS and MWU tests the density
distributions for all our SFR bins are shown to be different; for
SFR$>0-15\Msolar$\,yr$^{-1}$ the difference in environmental density
is present at the 2.6$\sigma$ level for both the KS-test and the MWU
test, for $15-30\Msolar$\,yr$^{-1}$ the KS and MWU tests shows a
difference at the 2.7$\sigma$ and 3.8$\sigma$ level, and for the
SFR$>30\Msolar$\,yr$^{-1}$ the distributions are significantly
different at the 3.3$\sigma$ and 4.8$\sigma$ levels, respectively.

This shows that the SFRs derived from the far-infrared emission of
galaxies are strongly linked to the environmental density of the
galaxy. Selecting only those galaxies with the highest SFRs results in
an even more pronounced difference between the normalized density
distributions of the Optical and FIR populations.

\subsection{Dust reddening effects}\label{sec:dustreddening}

\begin{table}
\raggedright
\caption{\label{tab:K2_tab} Two sample and two-dimensional KS and MWU-test results over the full SFR and redshift range ($0 < z \leq 0.5$). Where {\em{op}} represents Optical ($3,624$ objects) and {\em{FIR}} represents FIR ($1,208$ objects) matched in terms of their $K-$band magnitude, $r-K$ colour and redshift distribution. The two density distributions are different at the $> 5 \sigma$ level from KS tests, with the medians of the distributions different at the $> 5 \sigma$ level from MWU tests.}
\begin{tabular*}{8.44cm}{lcc}
  \hline
  Distributions Compared & KS Prob. & MWU Prob.\\
  \hline
  $z_{op}$ vs $z_{FIR}$ & 0.996 & 0.489 \\
    $(r-K)_{op}$ vs $(r-K)_{FIR}$ & 0.626 & 0.485 \\
  $m_{K(op)}$ vs $m_{K(FIR)}$ & 0.136 & 0.137 \\
  $(\bar{S}_{c})_{op}$ vs $(\bar{S}_{c})_{FIR}$ & $<10^{-9}$ & $<10^{-10}$ \\
  $(r-K, z)_{op}$ vs $(r-K, z)_{FIR}$ & 0.493 & -\\
  $(m_{K}, z)_{op}$ vs $(m_{K}, z)_{FIR}$ & 0.177 & - \\
  $(r-K, m_{K})_{op}$ vs $(r-K, m_{K})_{FIR}$ & 0.313 & -\\
  $(r-K, \bar{S}_{c})_{op}$ vs $(r-K, \bar{S}_{c})_{FIR}$ & $<10^{-5}$ & - \\
  $(m_{K}, \bar{S}_{c})_{op}$ vs $(m_{K}, \bar{S}_{c})_{FIR}$ & $<10^{-7}$ & - \\
  $(z, \bar{S}_{c})_{op}$ vs $(z, \bar{S}_{c})_{FIR}$ & $<10^{-5}$ & - \\  
  \hline
\end{tabular*}
\end{table}

In our analysis we have only used the most sensitive optical bands to
define our optical and far-infrared selected samples, due to the
wealth of such data over the survey region used. However, we would
expect that the galaxies which are detected in H-ATLAS to be subject
to dust reddening effects, as we know that they must have significant
amounts of dust in them to be detected in the first place. This could
cause our Optical and FIR galaxies to be mismatched in terms of their
intrinsic stellar colours and their stellar masses. To investigate
this we include the $K-$band data from the UKIRT Infrared Sky Survey
\citep{Lawrence_et_al2007}, which is available for many (but not all)
of our sources. We did not use this initially as the number of
galaxies with a $K-$band identification is less than the number which
are identified in the $g$, $r$ and $i$ bands. In
Figure~\ref{fig:rminK} we show a histogram of the $r-K$ colours of our
matched sample of Optical ({\em{red}}) and FIR ({\em{blue}}) galaxies
where, due to the smaller number of $K$-band detections, the size of
each sample is reduced to 2,139 and $759$ objects respectively. This
shows that there is indeed a significant difference in the $r-K$
distributions between the Optical and FIR galaxies at a $> 5\sigma$
level, suggesting that dust reddening may well be biasing our results.
However, this would only strengthen our results due to the fact that,
as we cross-match in the $r-$band, the reddening we see in
Figure~\ref{fig:rminK} is caused by the FIR population having brighter
$K-$band magnitudes than the Optical sample. Therefore, due to this
reddening, we are likely to be overestimating the FIR $K-$band
magnitudes and subsequently their masses. Given that we know more
massive galaxies generally trace denser environments, correcting for
this would lead to a larger difference between the Optical and FIR
samples.

To test the rigour of our result we repeat our density analysis with
Optical and FIR samples cross-matched in terms of $r-K$, $m_{K}$ and
$z$ parameter space. As this new cross-matching takes into
consideration only three dimensions, the resultant number of objects
considered matched in all three of these parameters is larger than in
our initial four-dimensional cross-matching, with $3,624$ Optical and
$1,208$ FIR objects.
Applying KS and MWU tests to the data return probability values
consistent with a significant difference between the environmental
density distributions to the $> 5\sigma$ level, with the FIR
population once more favouring underdense regions compared to the
Optical sample with mean values of $0.09 \pm 0.02$ and $0.23 \pm 0.01$
respectively. Table \ref{tab:K2_tab} gives the results of the
statistical comparison and the density distributions are plotted in
Figure \ref{fig:K2_histo}.

We do not extend on this analysis here as new data from the VISTA
VIKING Survey \citep[e.g.][]{Findlay_2012} over the full H-ATLAS
fields will mean that the analysis presented here could be carried out
with a $K-$band selected sample in the near future.

\subsection{Comparison with nearest-neighbour}\label{sec:NN_APP}
Applying the same KS and MWU statistical comparisons from Section
\ref{sec:KS-testing} between our cross-matched FIR and Optical data
sets, we find good agreement between the NN and the VT method; the two
NN defined normalized density distributions ($\bar{S}_{c}$) are
significantly different, with a KS test probability of $8.7 \times
10^{-5}$ indicating a significant difference at the $3.9\sigma$
level. As with our VT method all other parameter comparisons show no
significant difference as shown in Table \ref{tab:NNresults1}.  In
further agreement with the results established using the VT method,
the mean values of the two density distributions reveal that the
cross-matched FIR catalogue contains objects with lower environmental
densities than the Optical catalogue with mean values of $0.31 \pm
0.04$ and $0.60 \pm 0.05$ respectively. MWU tests reveal a significant
difference between the median values at the 4.6$\sigma$ level. In
addition, repeating the redshift binning from Section
\ref{sec:z-splitting} returns consistent results such that differences
are found in both bins increasing from 2$\sigma$ in the lowest bin to
3.3$\sigma$ in the higher bin.

\begin{table}
\raggedright
\caption{\label{tab:NNresults1} Two sample and two-dimensional KS and MWU-test results from the application of the 5th-nearest neighbour technique to the SDSS and H-ATLAS SDP data from Section \ref{sec:Analysis}. Where {\em{op}} represents Optical ($2,706$ objects) and {\em{FIR}} represents FIR ($902$ objects). The two density distributions are significantly different to the 3.9$\sigma$ level from KS tests in agreement with our VT technique.}
\begin{tabular*}{8.467cm}{lcc}
  \hline
  Distributions Compared & KS Prob. & MWU Prob.\\
  \hline
  $z_{op}$ vs $z_{FIR}$ & 0.999 & 0.403 \\
  $(g-r)_{op}$ vs $(g-r)_{FIR}$ & 0.974 & 0.374 \\
  $(r-i)_{op}$ vs $(r-i)_{FIR}$ & 0.395 & 0.239 \\
  $m_{r(op)}$ vs $m_{r(FIR)}$ & 0.719 & 0.290 \\
  $(\bar{S}_{c})_{op}$ vs $(\bar{S}_{c})_{FIR}$ & $<10^{-4}$ & $<10^{-5}$ \\
  $(g-r, r-i)_{op}$ vs $(g-r, r-i)_{FIR}$ & 0.223 & - \\
  $(g-r, z)_{op}$ vs $(g-r, z)_{FIR}$ & 0.956 & - \\
  $(r-i, z)_{op}$ vs $(r-i, z)_{FIR}$ & 0.468 & -\\
  $(m_{r}, z)_{op}$ vs $(m_{r}, z)_{FIR}$ & 0.755 & - \\
  $(g-r, m_{r})_{op}$ vs $(g-r, m_{r})_{FIR}$ & 0.839 & -  \\
  $(r-i, m_{r})_{op}$ vs $(r-i, m_{r})_{FIR}$ & 0.339 & -\\
  $(g-r, \bar{S}_{c})_{op}$ vs $(g-r, \bar{S}_{c})_{FIR}$ & 0.003 & - \\
  $(r-i, \bar{S}_{c})_{op}$ vs $(r-i, \bar{S}_{c})_{FIR}$ & 0.003 & - \\
  $(m_{r}, \bar{S}_{c})_{op}$ vs $(m_{r}, \bar{S}_{c})_{FIR}$ & 0.003 & - \\
  $(z, \bar{S}_{c})_{op}$ vs $(z, \bar{S}_{c})_{FIR}$ & 0.004 & - \\
  \hline
\end{tabular*}
\end{table}

\section{Comparison With Semi-Analytic Models}\label{sec:MillSim}

In this section we use the semi-analytic models (SAMs) of
\citet{Henriques_et_al2012}, who construct $24$ pencil-beam synthetic
light-cones for square areas (1.4 deg $\times$ 1.4 deg) out to high
redshift. These light-cones are based on the SAM of
\citet{Guo_et_al2011} which itself is built upon previous SAMs, such
as \citet{Croton_et_al2006}, to take into account a full range of
astrophysical processes; reionisation, cooling, disk size, star
formation, supernovae feedback, satellites, gas stripping, mergers,
bulge formation, black hole growth, feedback from active galactic
nuclei (AGN), metal enrichment, dust extinction, stellar mass and the
luminosity function. Each pencil-beam light-cone is constructed to
account for the fact that each side of the co-moving Millennium
Simulation box is equal to 500$h^{-1}$Mpc which is smaller than the
co-moving distance of an object at $z\sim2$. Therefore periodic
replication of the simulation can lead to multiples of the same object
being included due to discontinuities in large scale structure at the
boundaries between replications \citep{Kitzbichler&White2007}.

\begin{figure*}
\begin{center}
\includegraphics[width=0.693\columnwidth]{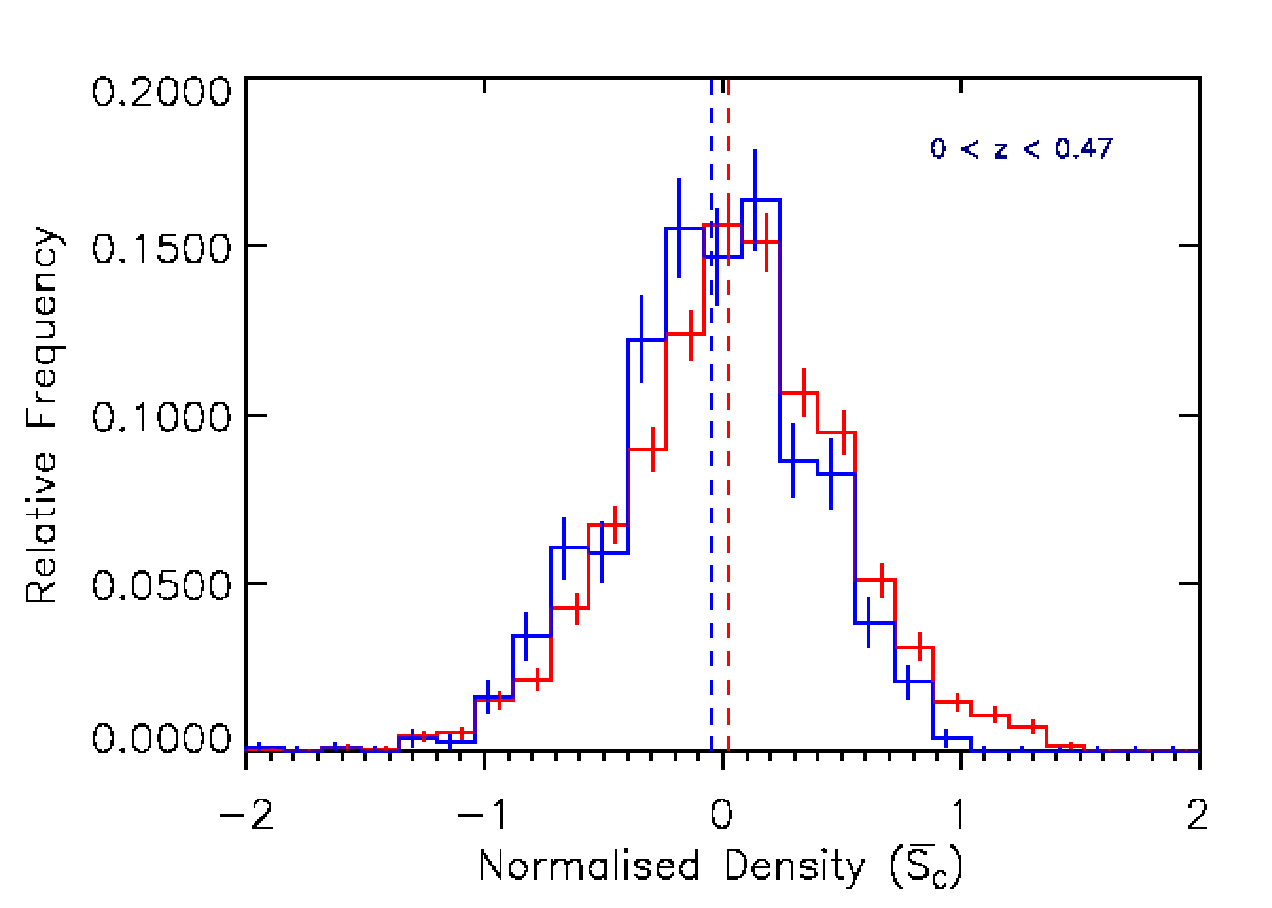}
\includegraphics[width=0.693\columnwidth]{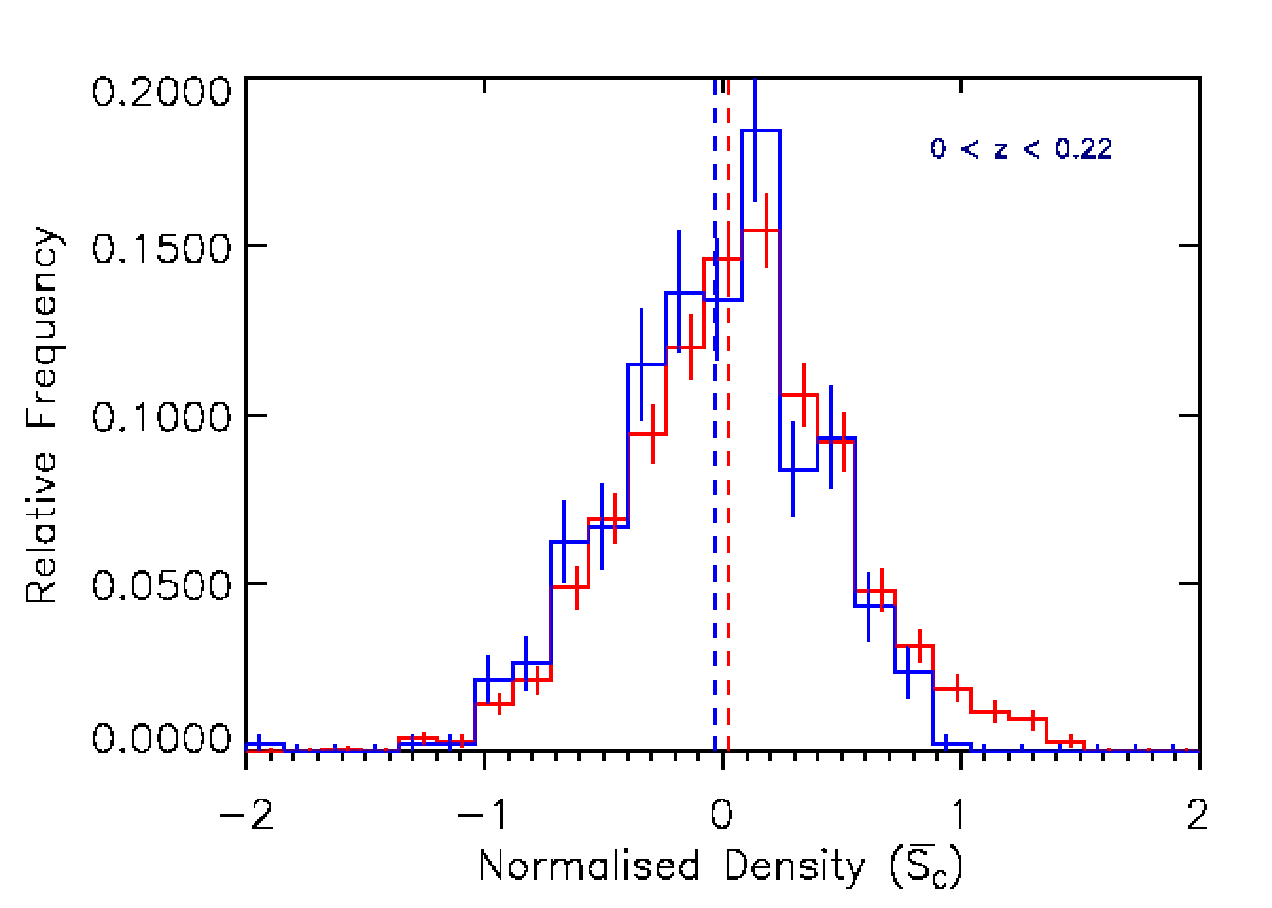}
\includegraphics[width=0.693\columnwidth]{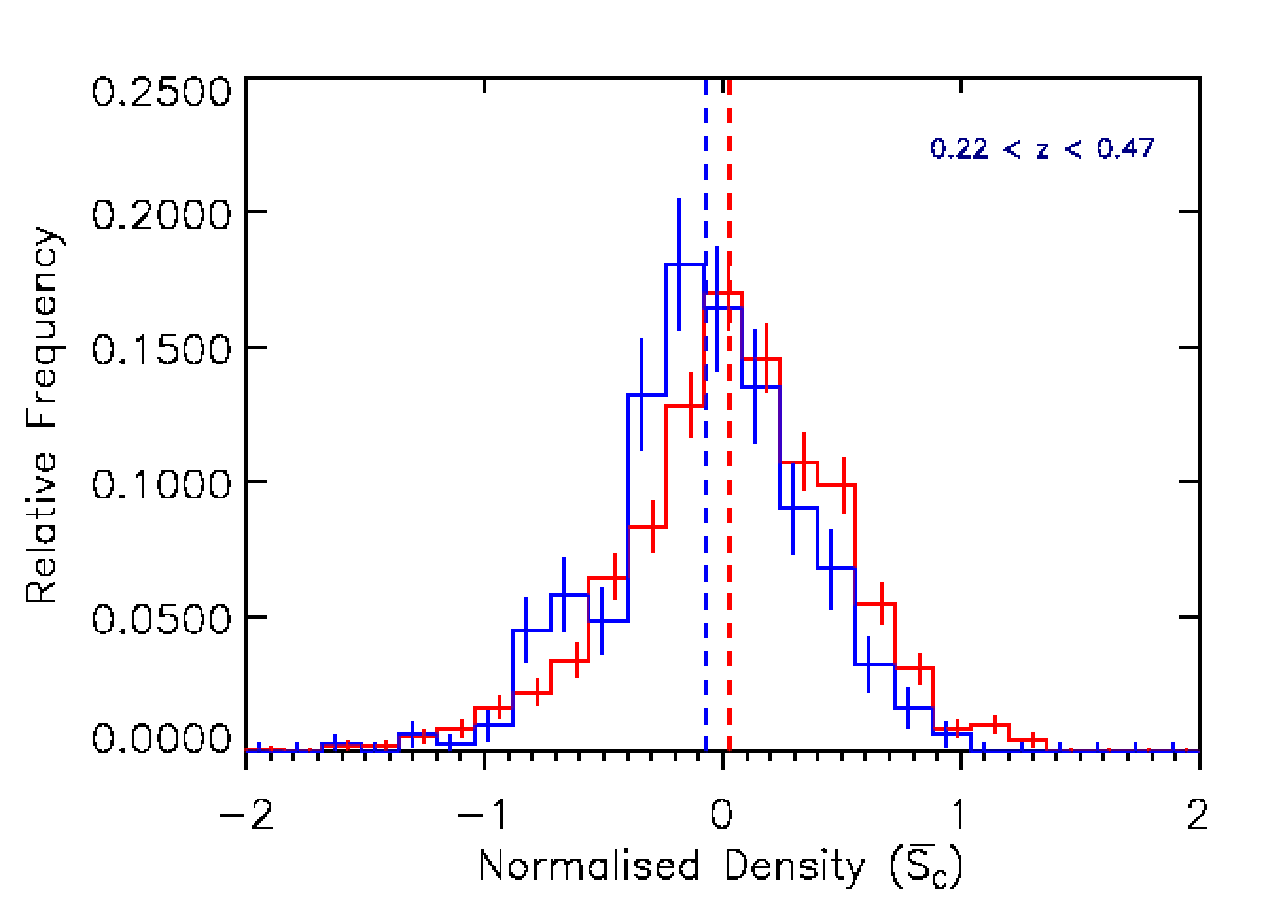}
\end{center}
\caption{A set of normalized histograms form the SAM output that show
  how the distribution of environmental density ($\bar{S}_{c}$) of the
  colour matched Optical-HG ({\em{red}}) and FIR-HG ({\em{blue}})
  populations change with redshift. {\em{Left}}: The full SFR range of
  both environmental density ($\bar{S}_{c}$) distributions ($2,184$
  and $728$ objects respectively), along with error bars depicting the
  normalized error associated with each bin. The two distributions are
  significantly different to the 4$\sigma$ level as determined by KS
  tests. {\em{Centre}}: The lower redshift bin ($0 < z \leq 0.22$) KS
  and MWU statistical tests reveal that the distributions are
  statistically different and that the null hypothesis can be rejected
  to at least the 2.4$\sigma$ level. {\em{Right}}: In the higher
  redshift bin ($0.22 < z \leq 0.47$) the distributions are again
  significantly different and the null hypothesis can be rejected to
  at least the 3.4$\sigma$ level.}
\label{fig:Ho4norm_HenGou}\label{fig:Hofullnorm_HenGuo}
\end{figure*}

\subsection{SAM analysis}\label{sec:Code App??}

Here we apply our environmental density measurement to the mock
catalogues of \citet{Henriques_et_al2012} in order to establish
whether the same relationship between environmental density and SFR is
found. Our results from Section \ref{sec:Analysis} have shown there is
a statistically significant difference between the density
distributions of galaxies with and without obscured star formation (as
traced by far-IR emission). This is such that galaxies with obscured
star formation favour underdense regions while galaxies without star
formation favour overdense regions.

We use all $24$ mock catalogues from \citet{Henriques_et_al2012}, with
additional data taken from the mock catalogue of
\citet{Guo_et_al2011}.  The data are selected to span the same
redshift range as our observed data ($0 < z \leq 0.5$), however these
catalogues do not initially contain any errors on their redshift
values. Therefore in order to achieve similar redshift sampling as
within our environmental density measurement in Section
\ref{sec:METHOD}, it is necessary to apply a redshift error to each
object based on matching the $r-$band magnitude ($m_{r}$) and
photometric redshift ($z$) values to the observed data catalogue. This
therefore establishes a likely value for the redshift error based on
these parameters. This is achieved by cross-matching each Henriques \&
Guo (hereafter HG) object to the total Optical-9hr catalogue, locating
all matches in $r-$band magnitude ($m_{r}$) and photometric redshift
($z$). For each match, the photometric error from the Optical-9hr
catalogue is applied as the redshift error to the HG object. Where a
spectroscopic redshift is located, a standard error for a
spectroscopic redshift ($0.001$) is applied, as in the initial
analysis (Section \ref{sec:METHOD}). In addition, as with the observed
data in Section \ref{sec:Optical Data}, an apparent $r-$band magnitude
cut was implemented removing all galaxies with magnitudes fainter than 21.5.
The resultant number of objects across the HG catalogue totals
$260,303$ objects. Applying our environmental density measure returns
normalized environmental density values in comparison to a random
field ($\bar{S}_{c}$) for each object, as with our observed data in
Section \ref{sec:Density Normalization}.  However, due to the smaller
field size ($-0.7<$RA$< 0.7$ degrees and $-0.7<$Dec$<0.7$ degrees) the
edge effect cuts imposed during the analysis have a greater impact on
the number of sources cut from our HG catalogue, further reducing the
catalogue to $112,125$ objects.

\begin{figure}
\centering
\includegraphics[width=1.0\columnwidth]{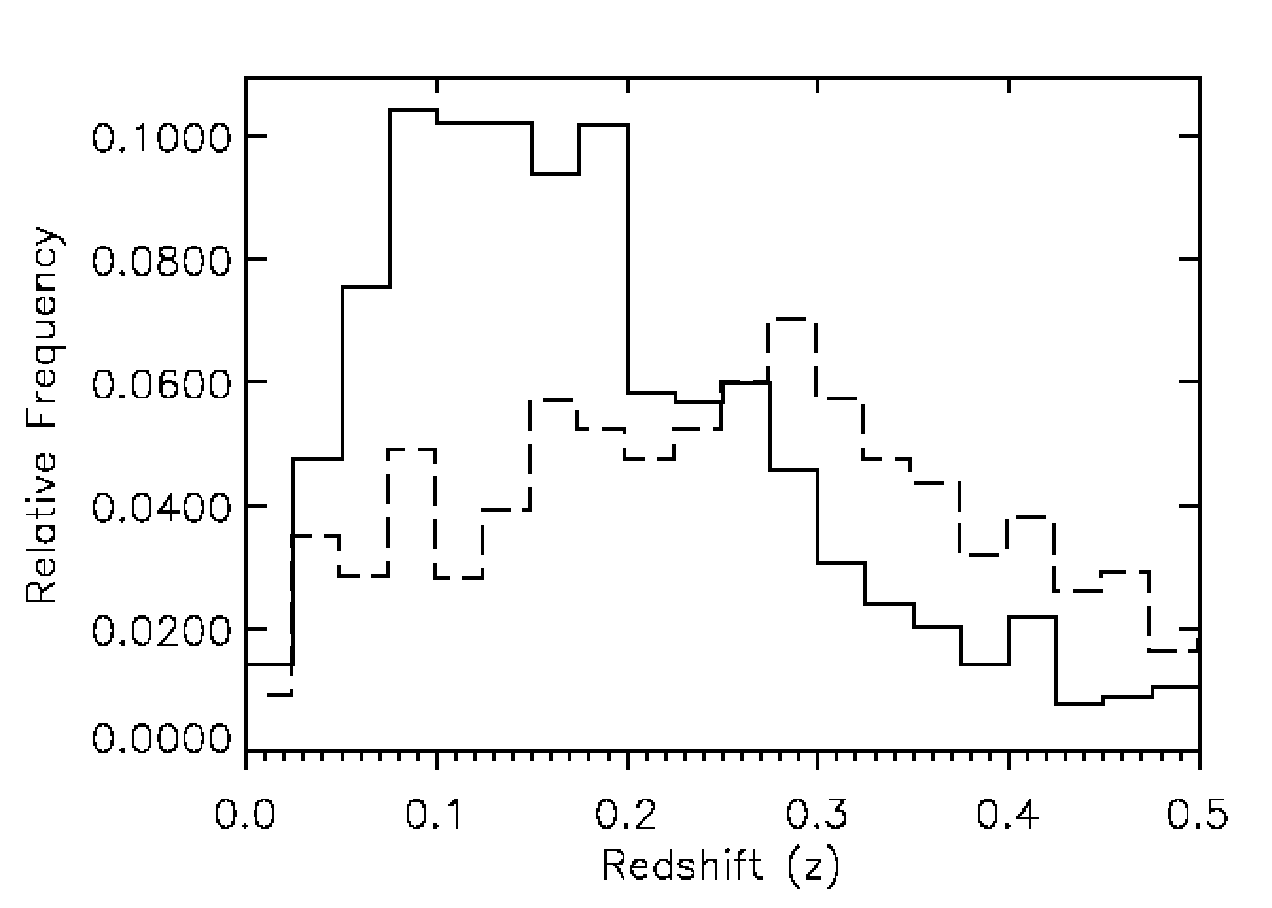}
\caption{Normalized histograms showing the comparison between the
  redshift distributions of both FIR and FIR-HG samples. {\em{Solid
      line:}} The total FIR-HG redshift distribution showing a clear
  weighting towards lower redshifts. {\em{Dashed line:}} The total FIR
  redshift distribution which peaks at higher redshifts.}
\label{fig:totalIRplots}
\end{figure}

For our analysis it is first necessary to determine which galaxies in
the simulated catalogue would have far-IR emission, and thus be
detectable by the H-ATLAS survey. This is achieved by calculating the
$250 \mu$m flux for each object from the given SFR and redshift values
given by the SAM. This is the direct reverse of the calculation in
Section \ref{sec:SFRlim} where we calculate the SFR, assuming a
temperature and emissivity index, from the $250 \mu$m flux of each FIR
galaxy. The average 5$\sigma$ $250 \mu$m flux limit of the H-ATLAS
observations ($33.5$ mJy), taken from \citet{Rigby_et_al2011},
provides an exact cut-off point for which a galaxy could be considered
detectable in the survey. From here the HG catalogue could be split
into IR and non-IR detected objects, equivalent to our FIR and Optical
catalogues in our observed data analysis from Section
\ref{sec:KS-testing0}. Objects with a $250$ $\mu$m flux density greater
than $33.5$ mJy are therefore considered detectable by H-ATLAS,
hereafter called FIR-HG ($1,919$ objects), and those with a $250$
$\mu$m flux density less than $33.5$ mJy are not considered detectable
by H-ATLAS and hereafter named Optical-HG ($110,206$
objects). 

As with our observed data in Section \ref{sec:KS-testing0}, a
like with like cross-matching process is applied.
We find $2,184$ objects from the Optical-HG population matched with
$728$ objects from the FIR-HG population. Both of these cross-matched
samples represent approximately the same percentage of their parent
populations as found with our observed cross-matched samples from
Section \ref{sec:KS-testing0}. This was such that the cross-matched
Optical and Optical-HG samples represent $\sim 2$ per cent of their
parent populations, with the FIR and FIR-HG samples representing $\sim
40$ and $\sim 38$\,per cent respectively.


\subsection{SAM statistical testing}\label{sec:KStesting2}

We perform the same statistical analysis, as with our observed data in
Section \ref{sec:KS-testing}, i.e. applying one- and two-dimensional
KS tests as well as MWU-tests to the two cross-matched populations,
the results of which are given in Table \ref{tab:MillResults}. These
values show, in agreement with our observed data in Section
\ref{sec:KS-testing}, that the null hypothesis can be rejected to at
least the 4$\sigma$ level for the normalized environmental density
($\bar{S}_{c}$) of the Optical-HG and FIR-HG populations. All other
parameter comparisons cannot be considered significantly different at
any reliable statistical level (e.g. $> 2 \sigma$).

The two density distributions are shown in
Figure~\ref{fig:Hofullnorm_HenGuo} ({\em{left}}). As expected, it is
the FIR-HG population ({\em{blue}}), with the mean of its distribution
at $\bar{S}_{c}=(-4.84 \pm 1.54)$ $\times 10^{-2}$ ({\em{blue dashed
    line}}), that is biased towards underdense regions, while the
Optical-HG population ({\em{red}}), with the mean of its distribution
at $\bar{S}_{c}=(2.59 \pm 1.72)$ $\times 10^{-2}$ ({\em{red dashed line}}),
shows a bias towards overdense regions. Errors included on each bin
are small and support the difference found between the
distributions. It is worth noting that if we do not include simulated
photometric errors and treat the SAM redshift values as precise, the
resultant $\bar{S}_{c}$ distributions exhibit a much larger spread in
values. Inclusion of photometric errors evidently reduces this spread
by essentially `washing out' the density structure we are trying to
recover. However, when precise redshifts values are used we find the
same correlations are found between the Optical-HG and FIR-HG
populations, despite the increased spread, with both populations
exhibiting a significant difference.

\begin{table}
\raggedright
\caption{\label{tab:MillResults} Full SFR range. Two sample and
  two-dimensional KS and MWU-test results over full redshift range ($0
  < z \leq 0.5$) where {\em{op}} represents Optical-HG ($2,184$
  objects) and {\em{FIR}} represents FIR-HG ($728$
  objects). The density distributions are significantly different to
  the 4$\sigma$ level from KS tests, with the median values of the distributions different to the 4.7$\sigma$ level from MWU tests:}
\begin{tabular*}{8.467cm}{lcc}
  \hline
  Distributions Compared & KS Prob. & MWU Prob.\\
  \hline
  $z_{op}$ vs $z_{FIR}$ & 0.658 & 0.318 \\
  $(g-r)_{op}$ vs $(g-r)_{FIR}$ & 0.990 & 0.452 \\
  $(r-i)_{op}$ vs $(r-i)_{FIR}$ & 0.198 & 0.172 \\
  $m_{r(op)}$ vs $m_{r(FIR)}$ & 0.237 & 0.125 \\
  $(\bar{S}_{c})_{op}$ vs $(\bar{S}_{c})_{FIR}$ & $<10^{-4}$ & $<10^{-5}$ \\
  $(g-r, r-i)_{op}$ vs $(g-r, r-i)_{FIR}$ & 0.131 & - \\
  $(g-r, z)_{op}$ vs $(g-r, z)_{FIR}$ & 0.751 & - \\
  $(r-i, z)_{op}$ vs $(r-i, z)_{FIR}$ & 0.278 & -\\
  $(m_{r}, z)_{op}$ vs $(m_{r}, z)_{FIR}$ & 0.081 & - \\
  $(g-r, m_{r})_{op}$ vs $(g-r, m_{r})_{FIR}$ & 0.343 & -  \\
  $(r-i, m_{r})_{op}$ vs $(r-i, m_{r})_{FIR}$ & 0.090 & -\\
  $(g-r, \bar{S}_{c})_{op}$ vs $(g-r, \bar{S}_{c})_{FIR}$ & $<10^{-3}$ & - \\
  $(r-i, \bar{S}_{c})_{op}$ vs $(r-i, \bar{S}_{c})_{FIR}$ & $<10^{-3}$ & - \\
  $(m_{r}, \bar{S}_{c})_{op}$ vs $(m_{r}, \bar{S}_{c})_{FIR}$ & 0.002 & - \\
  $(z, \bar{S}_{c})_{op}$ vs $(z, \bar{S}_{c})_{FIR}$ & 0.001 & - \\
  \hline
\end{tabular*}
\end{table}

\begin{table}
\centering
\caption{\label{tab:MilliResults2} Full SFR range KS-test and MWU-test results
  for both the Optical-HG and FIR-HG populations for
  $\bar{S}_{c}$ distributions within the individual redshift slices
  shown in Figure~\ref{fig:Ho4norm_HenGou}. The null hypothesis is
  rejected at the 2.4$\sigma$-2.9$\sigma$ level for both distributions in
  the lower redshift bin from KS and MWU tests. The null hypothesis is
  rejected at the 3.4$\sigma$-3.9$\sigma$ level in the higher redshift
  bin:}
\begin{tabular*}{8.38cm}{lcccc}
\hline
Redshift Slice & Optical & IR & KS Prob. & MWU Prob.\\
\hline
$0 \leq z < 0.22$ & 1,273 & 418 & 0.020 & 0.003\\ 
$0.22 \leq z < 0.47$ & 911 & 310 & $<10^{-3}$ & $<10^{-4}$ \\ 
\hline
\end{tabular*}
\end{table}

\begin{table*}
\raggedright
\caption{\label{tab:MillResults_5SFR} Two sample KS and MWU-test results where {\em{op}} represents Optical-HG and {\em{FIR}} represents
  FIR-HG. (A): SFR of 0 to 5 $\Msolar$\,yr$^{-1}$ containing
  Optical-HG $729$ objects and $243$ FIR-HG
  objects. The difference between the
  two density distributions in this bin cannot be distinguished. (B): SFR of 5 to 10 $\Msolar$\,yr$^{-1}$ containing $984$ Optical-HG objects and $328$
  FIR-HG objects. The density distributions are different
  at the 3.3$\sigma$ level from KS tests and 3.4$\sigma$ from MWU
  tests. (C): SFR of $>$ 10 $\Msolar$\,yr$^{-1}$ containing $669$
  Optical-HG objects and $223$ FIR-HG objects. From KS tests the two density distributions are different at the 3$\sigma$ level, with the median values of the distributions different at the 3.6$\sigma$ level from MWU tests:}
\begin{tabular*}{17.92cm}{lcccccc}
  \hline
  Distributions Compared & KS Prob. (A) & MWU Prob. (A) & KS Prob. (B) & MWU Prob. (B)  & KS Prob. (C) & MWU Prob. (C) \\
  \hline
  $z_{op}$ vs $z_{FIR}$ & 0.175 & 0.201 & 0.310 & 0.277 & 0.344 & 0.443\\
  $(g-r)_{op}$ vs $(g-r)_{FIR}$ & 0.947 & 0.328 & 0.965 & 0.423 & 0.770 & 0.322\\
  $(r-i)_{op}$ vs $(r-i)_{FIR}$ & 0.125 & 0.090 & 0.057 & 0.080 & 0.958 & 0.360\\
  $m_{r(op)}$ vs $m_{r(FIR)}$ & 0.065 & 0.134 & 0.909 & 0.301 & 0.216 & 0.089 \\
  $(\bar{S}_{c})_{op}$ vs $(\bar{S}_{c})_{FIR}$ & 0.450 & 0.098 & 0.001 & 0.001 & 0.004 & $<10^{-3}$\\
  \hline
\end{tabular*}
\end{table*}

To further examine the difference found between the Optical-HG and
FIR-HG populations we once more split the two populations into
redshift bins, repeating the analysis from Section
\ref{sec:z-splitting}. 
Figure~\ref{fig:totalIRplots} shows the redshift distributions of both
the FIR (dashed line) and FIR-HG (solid line) data. It is immediately
clear that the FIR-HG population is primarily weighted towards lower
redshifts with a mean value of 
($1.82 \pm 0.02$) $\times 10^{-1}$. Its number density falls off
beyond $z\sim0.25$ reaching a maximum redshift of $z=0.47$, short of
the full redshift range of the FIR sample. In comparison the FIR
population retains an approximately consistent number density across
its entire redshift range with a mean value
of 
($3.30 \pm 0.04$) $\times 10^{-1}$. Therefore it is necessary to
adjust the redshift binning from that applied to the FIR data in
Section \ref{sec:z-splitting} 
to account for this difference. Reducing the boundary between our
higher and lower redshift bins from $z=0.25$ to $z=0.22$, we bin the
Optical-HG and FIR-HG populations, re-plotting histograms to represent
the data within these redshift bins and reapplying KS-tests and MWU
tests to the data. The redshift bins are plotted into normalized
histograms displayed in Figure~\ref{fig:Ho4norm_HenGou} and the KS and
MWU test results from this redshift binning are given in Table
\ref{tab:MilliResults2}.  These statistical tests confirm that the
same SFR-density trend with redshift is found as with our
observational data analysis in Section
\ref{sec:z-splitting}.

This trend is found despite the differences between the redshift
distributions of both observed and simulated FIR data sets
(Figure~\ref{fig:totalIRplots}). We determine that a galaxy
constitutes part of the FIR-HG population by selecting objects based
on their $250 \mu$m flux, which is calculated from the individual SFR
derived from the SAM of \citet{Guo_et_al2011}. As noted in
Section~\ref{sec:SFRlim} our calculation of the SFR from the
far-infrared luminosity is subject to a range of assumptions that may
or may not be valid. Furthermore, the parameters used within the SAM
to calculate SFR may also not accurately incorporate all of the
physical processes that govern SFR. Detailed analyses of these effects
are beyond the scope of this paper, therefore we adopt a conservative
approach and consider that the results from the observations and the
SAMs are in qualitative agreement.

\subsection{Applying SFR limits to SAMs}\label{sec:SFRlimit2}

\begin{figure}
\centering
\includegraphics[width=1.0\columnwidth]{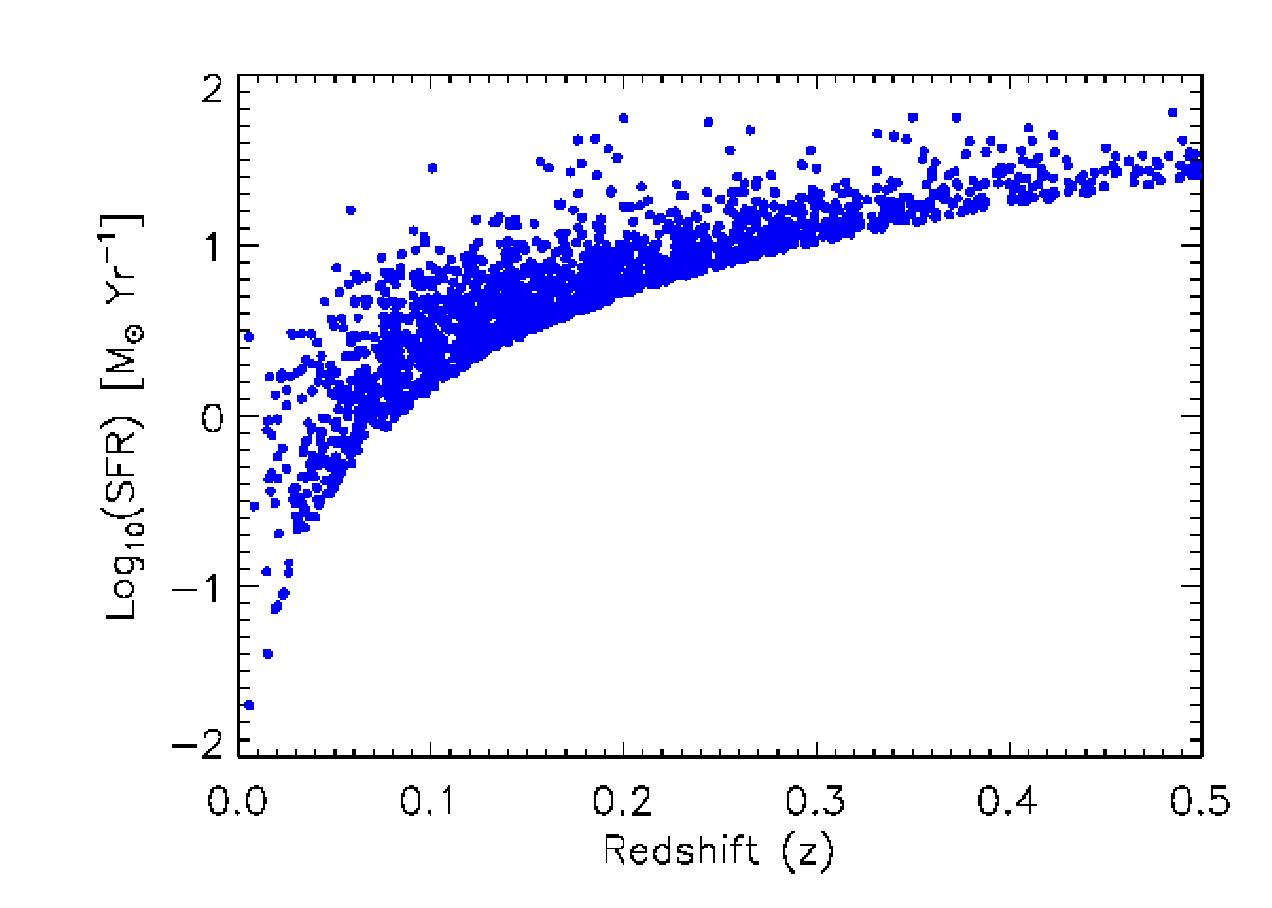}
\centering
\includegraphics[width=1.0\columnwidth]{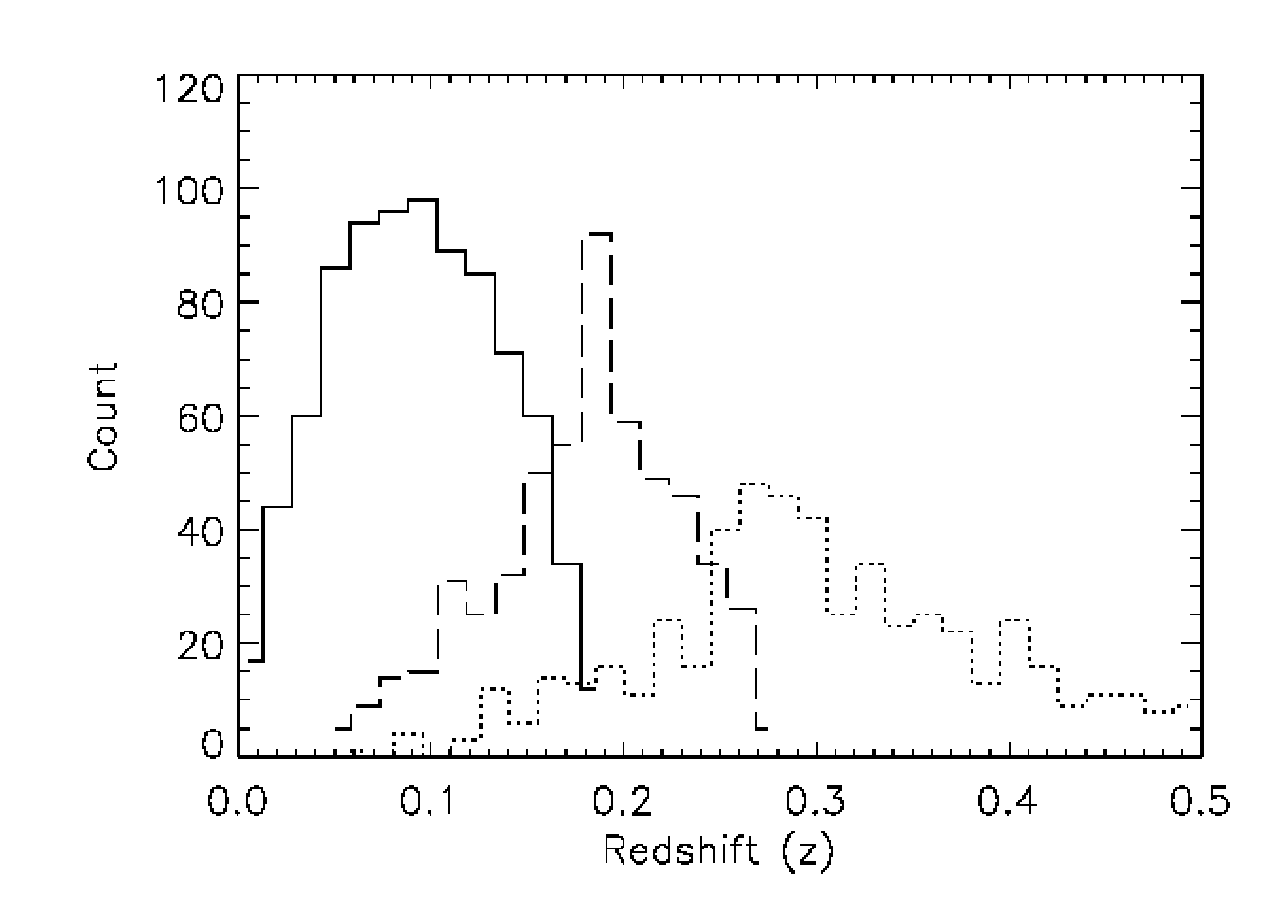}
\caption{{\em{Top}}: SFR ($\Msolar$\,yr$^{-1}$) versus
  redshift for the total FIR-HG catalogue. This shows that above 10 $\Msolar$\,yr$^{-1}$ the number of objects reduces
  significantly. {\em{Bottom}}: Plot of the three SFR bins from the
  FIR-HG catalogue vs redshift. The $0 < SFR < 5 \Msolar$\,yr$^{-1}$ bin
  (Solid line) containing $846$ objects, the $5 < SFR \leq 10 \Msolar$\,yr$^{-1}$ bin (Dashed line) containing $547$ objects and the $SFR >
  10 \Msolar$\,yr$^{-1}$ bin (Dotted line) containing $526$ objects.}
\label{fig:ZvsSFR_HenGuo}
\end{figure}

Here we repeat the same SFR binning analysis from Section
\ref{sec:SFRlim}. Applying this stage of the analysis to the
populations derived from SAMs allows us to further probe the
differences between the simulated data and the data obtained
observationally.  The reduction in the number of 
FIR-HG objects between the low and high redshift bins is illustrated
in Figure~\ref{fig:ZvsSFR_HenGuo}. This shows how the number density
of the FIR-HG population falls off beyond $z\sim0.25$, in comparison
with our observed data in Figure~\ref{fig:ZvsSFR}. It is evident that
the FIR-HG population has a higher fraction of sources at higher
redshifts than FIR population. Due to the vast majority of the FIR-HG
population residing below $z\sim0.25$ it is necessary to adjust the
SFR binning parameters, from those applied to the FIR population, to
narrower SFR ranges in order to achieve three comparative samples of
this population. We therefore bin the FIR-HG objects in terms of their
SFR in bins of $0-5$, $5-10$ and $>10 \Msolar$\,yr$^{-1}$. In
agreement with our results in Section \ref{sec:SFRlim}, we find that
with higher levels of star formation, the statistical difference
between the two populations from KS and MWU tests increases. Only in
our lowest SFR bin ($0-5 \Msolar$\,yr$^{-1}$) is no significant
difference found between the Optical-HG and FIR-HG samples. These
values are presented in Table \ref{tab:MillResults_5SFR}.



Figure~\ref{fig:ZvsSFR_HenGuo} shows how at higher SFRs the number
density of objects reduces to such an extent that it makes further
analysis unfeasible. Therefore it was not possible to introduce SFR
bins at higher SFRs than 10 $\Msolar$\,yr$^{-1}$ to the
analysis. Despite this, some key conclusions can be made with regards
to the Optical-HG/ FIR-HG comparison based on the three SFR bins
applied to the data. This analysis has shown that, for galaxies with
SFRs higher than 5 $\Msolar$\,yr$^{-1}$, there is a statistically
significant difference between both of the $\bar{S}_{c}$ distributions
and that this statistical difference becomes more pronounced in these
higher SFR bins as a result of removing lower star-forming objects
from the comparison. From finding no statistical difference between
the density distributions of the Optical-HG and FIR-HG populations
when SFRs are less than 5 $\Msolar$\,yr$^{-1}$ to finding a
significant difference to at least the 3$\sigma$ level in higher SFR
bins.

\subsection{Application of NN to SAMs}\label{sec:NN_SAM}

We again test the $N$th-nearest neighbour against our Voronoi
Tessellation methods. We apply the NN method to our analysis of
semi-analytic models. Following the same processes from Sections
\ref{sec:Code App??}  and \ref{sec:KStesting2} we apply our algorithm,
changed to incorporate the NN method, to the total HG catalogue, again
dividing the output according to the H-ATLAS flux limit and
cross-matching in $g-r$, $r-i$, $z$ and $m_{r}$ parameter space. This,
once more, provides two catalogues representative of optical and
far-infrared (Optical-HG and FIR-HG) that can be accurately compared
to analyse differences in density.

We find that our results match those of Section~\ref{sec:KStesting2},
with KS and MWU test results finding a significant difference between
the normalized densities of the Optical-HG and FIR-HG populations only
(Table \ref{tab:NNresults2}). With a KS-test and MWU-test
probabilities of $\sim 10^{-5}$ indicating a significant difference at
the $\gtsim 4\sigma$ level in both cases.  The mean values of each
distribution lie at ($4.02 \pm 0.24$) $\times 10^{-1}$ for the
Optical-HG and ($1.60 \pm 0.21$) $\times 10^{-1}$ for the FIR-HG
indicating that the Optical-HG population occupy generally more
overdense regions in agreement with our study using VT.

\begin{table}
\raggedright
\caption{\label{tab:NNresults2} Two sample and two-dimensional KS and MWU-test
  results from the application of the 5th-nearest neighbour technique to
  our semi-analytic model analysis Optical-HG and FIR-HG
  populations. Where {\em{op}} represents Optical-HG ($2,184$
  objects) and {\em{FIR}} represents FIR-HG ($728$ objects). The
  two density distributions are significantly different to the
  4.5$\sigma$ level from KS tests in agreement with our VT technique.}
\begin{tabular*}{8.467cm}{lcc}
  \hline
  Distributions Compared & KS Prob. & MWU Prob.\\
  \hline
  $z_{op}$ vs $z_{FIR}$ & 0.658 & 0.318 \\
  $(g-r)_{op}$ vs $(g-r)_{FIR}$ & 0.990 & 0.452 \\
  $(r-i)_{op}$ vs $(r-i)_{FIR}$ & 0.198 & 0.172 \\
  $m_{r(op)}$ vs $m_{r(FIR)}$ & 0.237 & 0.125 \\
  $(\bar{S}_{c})_{op}$ vs $(\bar{S}_{c})_{FIR}$ & $<10^{-4}$ & $<10^{-5}$ \\
  $(g-r, r-i)_{op}$ vs $(g-r, r-i)_{FIR}$ & 0.131 & - \\
  $(g-r, z)_{op}$ vs $(g-r, z)_{FIR}$ & 0.751 & - \\
  $(r-i, z)_{op}$ vs $(r-i, z)_{FIR}$ & 0.278 & -\\
  $(m_{r}, z)_{op}$ vs $(m_{r}, z)_{FIR}$ & 0.081 & - \\
  $(g-r, m_{r})_{op}$ vs $(g-r, m_{r})_{FIR}$ & 0.343 & -  \\
  $(r-i, m_{r})_{op}$ vs $(r-i, m_{r})_{FIR}$ & 0.090 & -\\
  $(g-r, \bar{S}_{c})_{op}$ vs $(g-r, \bar{S}_{c})_{FIR}$ & $<10^{-3}$ & - \\
  $(r-i, \bar{S}_{c})_{op}$ vs $(r-i, \bar{S}_{c})_{FIR}$ & $<10^{-3}$ & - \\
  $(m_{r}, \bar{S}_{c})_{op}$ vs $(m_{r}, \bar{S}_{c})_{FIR}$ & 0.001 & - \\
  $(z, \bar{S}_{c})_{op}$ vs $(z, \bar{S}_{c})_{FIR}$ & $<10^{-3}$ & - \\
  \hline
\end{tabular*}
\end{table}

\section{Discussion}\label{sec:Discussion}

The increased statistical separation between the Optical and FIR
density distributions, found with both increasing redshift and
increasing SFR, provides a clue to the role of environment in the
evolution of galaxies over the redshift range ($0 < z \leq 0.5$).
We find a clear segregation in the galaxy environmental density
between far--infrared-detected sources and those galaxies that are
matched in terms of optical colour, magnitude and redshift but devoid
of detectable far-infrared emission. Moreover, we find that this
segregation becomes more pronounced at brighter far-infrared
luminosity (or SFR) or at higher redshift. Unfortunately our data
precludes us from distinguishing between an evolutionary effect and
one associated with the level of star formation activity.

It is important to note that the reliability criterion ($R>0.8$), that
we employ from \citet{Smith_et_al2011} to select optical counterparts
to the FIR data, does not present a bias in our results. This
potential bias is such that in denser regions, with an increased
number of potential optical counterparts to a FIR object, the
reliability parameter for that FIR object, as defined in
\citet{Smith_et_al2012}, would reduce. In other words, in denser
regions it potentially becomes more difficult to reliably associate
the FIR detection with a unique optical source. Therefore the FIR
object may be excluded leaving only the FIR detections in relatively
low-density environments.
We tested this bias by making a more inclusive cut to the FIR sample
based on the minimum likelihood-ratio (LR) rather than the reliability
criterion (R). Our FIR sample was therefore increased to include these
previously missing sources. Upon repeating the analysis we found that
we still obtained similarly significant differences between the FIR
and Optical samples.

These results support previous studies that also suggest that the
presence of star formation in a galaxy is negatively correlated with
the density of its environment (e.g., \citealt{Dressler1980};
\citealt{Postman&Geller1984}; \citealt{Dressler_et_al1997};
\citealt{Dominguez_et_al2001}; \citealt{Goto_et_al2003};
\citealt{Kauffmann_et_al2004}; \citealt{O'Mill_et_al2008};
\citealt{Lee_et_al2010}). Our analysis has shown that this correlation
holds on individual galaxy scales, and thus the processes responsible
for this correlation must have influence at this level as well as on
larger scales. In addition, our use of far-infrared observations mean
our results are not affected by uncertainties associated with
extinction or H$\alpha$ to SFR conversions.

However, the exact mechanism responsible for the observed reduction of
SFR with increase in density remains uncertain.
Recent studies by \citet{Deng_et_al2011} and
\citet{Wijesinghe_et_al2012} suggest that there is no trend with
environment when restricting the SFR-density comparison to purely
star-forming objects. They conclude, therefore, that the observed
SFR-density correlation is due to the increasing fraction of passive
galaxies across the total galaxy sample since
$z\sim1$. \citet{Deng_et_al2011} go further and suggest that the
SFR-density relation is strongly colour dependent, with blue galaxies
exhibiting a very weak correlation between environment and SFR. In
contrast, they find red galaxies to exhibit strong correlation between
environmental density and SFR attributing this to the increasing
presence of red late-type morphologies. As our analysis has focused on
the direct comparison of star formation properties with the individual
environmental densities of each object, we have shown that there is a
clear difference between the star-forming and passive population in
our colour-matched samples in agreement with earlier work by
\citet{Gomez_et_al2003} and \citet{Welikala_et_al2008}.

Furthermore, we have carried out the same analysis on SAMs where we
obtain a similar result, i.e. the environmental density distributions
from the total simulated Optical-HG and FIR-HG populations were found
to be significantly different at the 4$\sigma$ level. Qualitative
agreement is also found when we bin in terms of both redshift and
SFR. 


\section{Summary \& Conclusions}\label{sec:Conclusions}

We have compared the environmental and star formation properties of
two populations of galaxies out to $z \sim0.5$. For this analysis we
have used optical spectroscopy and photometry from the GAMA 9hr survey
(DR1 data) and SDSS, with far-infrared observations from the H-ATLAS
SDP. We use Voronoi Tessellations to analyse the environmental
densities of these galaxies on individual scales normalized to account
for differences in the population density and uniformity across the
redshift range due to
the flux limit of the survey and the increasing volume sampled with
increasing redshift.

The environmental density of the Optical and far-IR catalogues were
then compared by initially matching the catalogues in
multi-dimensional colour, magnitude and redshift space ($g-r$, $r-i$,
$m_{r}$, $z$) selecting a matched population of the Optical sources
numbering three times that of the far-IR distribution, in order to
obtain a robust comparison over $0 < z \leq 0.5$. Our key results are:

\begin{enumerate}
\item Objects with far-IR detected emission, and levels
of star formation $>5\Msolar$\,yr$^{-1}$, reside in less dense
environments than galaxies not detected at far-infrared
wavelengths.

\item The environmental density difference between the
two far-IR and non-far-IR luminous galaxies also increases with
redshift, with a 2.2$\sigma$ difference in the lower bin ($0 < z \leq
0.25$) and a 3.3$\sigma$ difference in the higher bin ($0.25 < z \leq
0.50$), with the far-infrared detected galaxies again residing in less
dense environments. In relation to this, we find an increasing
separation between the density distributions with increasing SFR from
2.6$\sigma$, 2.7$\sigma$ and 3.3$\sigma$ respectively, although we
note that we cannot distinguish redshift effects from luminosity
effects in our flux-density limited sample.


\item We find substantial differences between redshift
distributions of both our observed and SAM far-infrared samples. This
provides interesting indications on how recipes for star formation
need to be modified within SAMs to improve their ability to model the
observed universe.

\item We also note that VT are a reliable and accurate method
of calculating the environmental densities for individual
galaxies. Indeed, the use of VT for this purpose may surpass the NN
technique, as their improved resolution is able to measure more detailed
density structure.

\end{enumerate}

\section{Acknowledgements}\label{sec:Acknowledge}

The {\it Herschel}-ATLAS is a project with {\it Herschel}, which is an
ESA space observatory with science instruments provided by
European-led Principal Investigator consortia and with important
participation from NASA. The H-ATLAS website is:\\
http://www.h-atlas.org/\\
GAMA is a joint European-Australasian project based around a
spectroscopic campaign using the Anglo- Australian Telescope. The GAMA
input catalogue is based on data taken from the Sloan Digital Sky
Survey and the UKIRT Infrared Deep Sky Survey. Complementary imaging
of the GAMA regions is being obtained by a number of independent
survey programs including GALEX MIS, VST KIDS, VISTA VIKING, WISE,
Herschel-ATLAS, GMRT and ASKAP providing UV to radio coverage. GAMA is
funded by the STFC (UK), the ARC (Australia), the AAO, and the
participating institutions. The GAMA website is:
http://www.gama-survey.org/

\bibliography{bib2}{} \bibliographystyle{mn2e} 

\end{document}